\documentclass{aa}
\usepackage[varg]{txfonts}
\usepackage[utf8]{inputenc}
\usepackage{wasysym}
\usepackage{natbib}
\usepackage{graphicx}
\usepackage{gensymb}
\usepackage{textcomp}
\usepackage{hyperref}
\usepackage{booktabs}
\usepackage{enumitem}
\usepackage{amsmath}

\bibpunct{(}{)}{;}{a}{}{,}

\begin{document}

\title{The multifarious ionization sources and disturbed kinematics of extraplanar gas in five low-mass galaxies}

\author{R.~P.~V.~Rautio \inst{\ref{inst1}}
  \and A.~E.~Watkins \inst{\ref{inst2}}
  \and S.~Comer\'on \inst{\ref{inst3},\ref{inst4}}
  \and H.~Salo \inst{\ref{inst1}}
  \and S.~D\'iaz-Garc\'ia \inst{\ref{inst3},\ref{inst4},\ref{inst5}}
  \and J.~Janz \inst{\ref{inst6},\ref{inst1},\ref{inst7}}}

\institute{Space Physics and Astronomy research unit, University of Oulu, 90014 Oulu, Finland \\\email{riku.rautio93@gmail.com} \label{inst1}
  \and Centre of Astrophysics Research, School of Physics, Astronomy and Mathematics, University of Hertfordshire, Hatfield AL10 9AB, UK \label{inst2}
  \and Departamento de Astrof\'isica, Universidad de La Laguna, 38200 La Laguna, Tenerife, Spain \label{inst3}
  \and Instituto de Astrof\'isica de Canarias, 38205 La Laguna, Tenerife, Spain \label{inst4}
  \and Department of Physics, Engineering Physics and Astrophysics, Queen's University, Kingston, ON K7L 3N6, Canada \label{inst5}
  \and Finnish Centre of Astronomy with ESO (FINCA), Vesilinnantie 5, FI-20014 University of Turku, Finland \label{inst6}
  \and Specim, Spectral Imaging Ltd., Elektroniikkatie 13, FI-90590 Oulu, Finland \label{inst7}}

\abstract{}{We investigate the origin of the extraplanar diffuse ionized gas (eDIG) and its predominant ionization mechanisms in five nearby (17--46 Mpc) low-mass ($10^9\text{--}10^{10}$ $M_{\odot}$) edge-on disk galaxies: ESO~157-49, ESO~469-15, ESO~544-27, IC~217, and IC~1553.}
         {We acquired Multi Unit Spectroscopic Explorer (MUSE) integral field spectroscopy and deep narrowband H$\alpha$ imaging of our sample galaxies. To investigate the connection between in-plane star formation and eDIG, we measure the star formation rates (SFRs) and perform a photometric analysis of our narrowband H$\alpha$ imaging. Using our MUSE data, we investigate the origin of eDIG via kinematics, specifically the rotation velocity lags. We also construct standard diagnostic diagrams and emission-line maps (EW(H$\alpha$), [N{~\sc ii}]/H$\alpha$, [S{~\sc ii}]//H$\alpha$, [O{~\sc iii}]/H$\beta$) and search for regions consistent with ionization by hot low-mass evolved stars (HOLMES) and shocks.}
         {We measure eDIG scale heights of $h_{z\text{eDIG}} = 0.59 \text{--} 1.39$ kpc and find a positive correlation between them and specific SFRs. In all galaxies, we also find a strong correlation between extraplanar and midplane radial H$\alpha$ profiles. These correlations along with diagnostic diagrams suggest that OB stars are the primary driver of eDIG ionization. However, we find regions consistent with mixed OB--HOLMES and OB--shock ionization in all galaxies and conclude that both HOLMES and shocks may locally contribute to the ionization of eDIG to a significant degree. From H$\alpha$ kinematics, we find rotation velocity lags above the midplane with values between 10 and 27 km s$^{-1}$ kpc$^{-1}$. While we do find hints of an accretion origin for the ionized gas in ESO~157-49, IC~217, and IC~1553, overall the ionized gas kinematics of our galaxies do not match a steady galaxy model or any simplistic model of accretion or internal origin for the gas.}
         {Despite our galaxies' similar structures and masses, our results support a surprisingly composite image of ionization mechanisms and a multifarious origin for the eDIG. Given this diversity, a complete understanding of eDIG will require larger samples and composite models that take many different ionization and formation mechanisms into account.}


\maketitle

\section{Introduction}
Diffuse ionized gas (DIG) is a component of the interstellar medium (ISM), the non-stellar baryonic component of galaxies. This extensive, warm ($10^4$ K), low density ($10^{-1}$ cm$^{-3}$) gas, composed primarily of ionized hydrogen, was discovered in our own Galaxy more than five decades ago \citep{hoyle1963ionized} and in other galaxies in the early 1990s \citep{dettmar1990dig, rand1990wim}. In the Milky Way, the DIG is often called the warm ionized medium (WIM) or the Reynolds Layer and has a scale height of 1.0--1.8 kpc (e.g., \citealt{haffner1999sh, gaensler2008sh}). While $ \sim 20 \% $ of the Galactic gas overall is ionized, the WIM fraction reaches nearly $100\%$ at heights above the midplane greater than a few hundred parsecs \citep{ferriere2001ism}.

The source of this ionization is thought mainly to be young, luminous O and B stars (e.g., \citealt{hoyle1963ionized, rand1996diginedge, levy2019edge}). Near these hot stars, gas is photoionized up to a distance where the density of the Lyman continuum (LyC) photons drops enough that recombination and ionization balance out, giving birth to H{~\sc ii} regions. To ionize the DIG, the H{~\sc ii} regions must be ``leaky,'' with either empty holes or lower density sections to allow the LyC photons to escape to the surrounding ISM \citep{haffner2009leaky, weber2019leaky}. However, DIG has also been found outside the optically luminous disks of galaxies, both at large projected distances (e.g., \citealt{williams1993virgo, lehnert1999m82, devine1999m82, kenney2008m86, lintott2009voorwerp, boselli2016ngc4569, watkins2018m51}) and at great extraplanar heights (e.g., \citealt{rand1996diginedge, miller2003twodisk, jo2018fuv, levy2019edge}), kiloparsecs away from the in-plane H{~\sc ii} regions. For the ionizing radiation to be able to propagate to these distances, not only does the radiation have to escape the H{~\sc ii} regions, but it must somehow find absorption-free pathways through the opaque clouds of neutral atomic hydrogen that are prevalent near the midplane.

Another observed feature of DIG that is hard to explain through OB star photoionization is its emission-line spectrum. Many emission-line ratios observed in the DIG (i.e., [O{~\sc iii}]$_{\lambda 5007}$/H$\beta$, [N{~\sc ii}]$_{\lambda 6583}$/H$\alpha$, [S{~\sc ii}]$_{\lambda 6716}$/H$\alpha$, and [O{~\sc i}]$_{\lambda 6300}$/H$\alpha$) are enhanced when compared to those observed in H{~\sc ii} regions (e.g., \citealt{rand1997digspec1,rand1998digspec2, collins2001dig, seon2009lyc}). The line ratios in a gas recombination emission-line spectrum are affected by the temperature and density of the gas, as well as by the spectrum of the ionizing radiation. Harder, or more energetic, photons may ionize species with higher ionization potentials than softer, or less energetic, photons. If the DIG is ionized by radiation escaping through holes in the H{~\sc ii} regions, the spectrum of the ionizing radiation will be that of the OB star, and one would expect to find similar emission-line spectra in the DIG as in the H{~\sc ii} regions. While some of the differences between the DIG and H{~\sc ii} region emission-line spectra (i.e., enhanced [N{~\sc ii}]/H$\alpha$ and [S{~\sc ii}]/H$\alpha$) can be explained by the differences in temperatures and densities, others (i.e., enhanced [O{~\sc iii}]/H$\beta$ and [O{~\sc i}]/H$\alpha$) cannot (e.g., \citealt{haffner2009leaky}). Spectral processing in the H{~\sc ii} regions or the ISM beyond them, such as the hardening of the ionizing radiation due to partial absorption, can further explain some of the observed line ratios in DIG, but most photoionization models still struggle to reproduce the DIG emission-line spectrum fully \citep{hoopes2003hard, wood2004hard, wood2005hard, haffner2009leaky}.

Several other mechanisms have been proposed to complement the photoionization by OB stars. Evolved stars such as post asymptotic giant branch stars and white dwarfs can have much greater scale heights than the in-plane OB stars, and they can in principle produce the emission-line ratios prevalent in the DIG \citep{flores-fajardo2011holmes, alberts2011evolution, zhang2017manga, lacerda2018ew}. Galaxy tidal interactions and feedback mechanisms such as those induced by supernova explosions, stellar winds, and active galactic nucleus (AGN) jets can cause shock ionization (e.g., \citealt{dopita1995shocks, collins2001dig, simpson2007shock, rich2011shocks, ho2014shocks, molina2018liner}). Interfaces between hot and cool gas \citep{cowie1977evaporation,begelman1990tml,shapiro1991fountain,slavin1993turbulent}, photoelectric heating from interstellar dust particles or large molecules \citep{reynolds1992wim, weingartner2001photoelectric}, dissipation of turbulence \citep{minter1997dissipation}, and magnetic reconnection \citep{raymond1992microflare} have also been proposed. Additionally, a fraction of the extraplanar H$\alpha$ emission could be light from the midplane scattered by extraplanar dust \citep{ferrara1996scatter, wood1999scatter, jo2018fuv}. Many mechanisms are most likely at play in real galaxies, and their relative importance can be expected to depend on the properties of both the host galaxy and the DIG. Deciphering these correlations is crucial for understanding the DIG.

Of special interest is the extraplanar DIG (eDIG). It is located at the interface between the disk and the halo components of galaxies, where important feedback mechanisms such as gas outflows driven by superbubbles are expected to take place. For example, simulations have shown that superbubble-driven outflows can regulate star formation (SF) in low-mass galaxies \citep{keller2016sb}, while eDIG has been shown to be nearly ubiquitous in late-type edge-on galaxies \citep{miller2003twodisk, bizyaev2017manga, levy2019edge}. This dynamic nature of eDIG is also reflected in its morphology, which varies from fairly smooth to heavily disturbed and turbulent with filaments extending up to several kiloparsecs (e.g., \citealt{dettmar1990dig, rand1990wim, collins2000dig, miller2003twodisk}). The morphology also varies spectrally depending on the ionization source, and emission-line maps can be used to trace the outflows (e.g., \citealt{lopez-coba2019outflows}). The eDIG is connected to both SF via leaky H{~\sc ii} regions and to the baryonic feedback mechanisms that are required by modern cosmological simulations to regulate SF and produce realistic galaxies (e.g., \citealt{naab2017feedback}). Therefore, studying eDIG in low-mass galaxies can give us valuable insights into the energetic connection between the disk and halo components of galaxies and can help bridge the gap between cosmological models and small-scale observations.

To do this, special consideration must be given to the observations, as the eDIG is a low-surface-brightness feature and mapping it requires reaching an H$\alpha$ surface brightness level lower than $\sim 10^{-16}$ erg s$^{-1}$ cm$^{-2}$ arcsec$^{-2}$ (i.e., it has emission measure below $\sim 50$ cm$^{-6}$ pc). In addition to good sensitivity, a comprehensive study of eDIG requires good spatial resolution, as well as the coverage of several emission lines. For this reason, integral field unit (IFU) spectroscopy has emerged as an excellent tool in its study. Recent IFU spectroscopy has demonstrated the complex kinematics and emission-line properties inherent to the eDIG with unprecedented detail \citep{jones2017manga, levy2019edge}. Unfortunately, current IFU spectrographs have limited fields of view (FoVs), and they are oversubscribed given their multifaceted uses. They also require a substantial amount of observation time to reach the low surface brightness necessary for eDIG observations. Narrowband emission-line imaging provides a faster, albeit less detailed, alternative, with a vast selection of different sized telescopes and instruments with different FoVs available. While it is limited to targeting only one band at a time, this is not an issue for morphological studies, and narrowband imaging is still an indispensable method to complement IFU observations. 

To fully describe the eDIG, in addition to understanding its morphology and emission-line properties, one must also understand its kinematics. The kinematics are especially useful in determining the origin of the extraplanar gas, which may be external -- accretion from the intergalactic or circumgalactic medium \citep{oort1970acc, binney2005accretion, kaufmann2006accretion, sancisi2008accretion} -- or internal -- ejected from the galaxy via galactic fountains \citep{shapiro1976ism,bregman1980fountain, fraternali2002fountain, fraternali2008fountain, fraternali2015fountain}. Radial variations in the negative vertical velocity gradient, or lag, of the ionized gas are expected to depend on the origin of the extraplanar gas \citep{levy2019edge}. \cite{bizyaev2017manga} argued that the correlations they found between the lag and the galactic stellar mass, central velocity dispersion, and axial ratio of the light distribution suggest a possible higher ratio of infalling-to-local gas in massive early-type disk galaxies compared to lower-mass late-type galaxies.

In this work we present the first detailed study of eDIG in a sample of nearby low-mass edge-on galaxies. Our sample is drawn from the sample of \cite{comeron2019thick}, who serendipitously discovered eDIG in Multi Unit Spectroscopic Explorer (MUSE; \citealt{bacon2010muse}) data of eight edge-on galaxies drawn from the \emph{Spitzer} Survey of Stellar Structure in Galaxies (S$^4$G; \citealt{sheth2010s4g}). To obtain a complete picture of the extended H$\alpha$ morphology, we acquired deep H$\alpha$ narrowband imaging of five of those galaxies with the New Technology Telescope (NTT) at La Silla Observatory and the Nordic Optical Telescope (NOT) at the Observatorio del Roque de los Muchachos. Our sample offers a unique combination of depth, spatial resolution, and wavelength coverage for the study of eDIG ionization and formation mechanisms in low-mass galaxies. In our analysis, we use H$\alpha$ profiles and emission-line diagnostics to investigate the prevalent ionization mechanisms for the extraplanar gas, and ionized gas kinematics to investigate the origin of the gas.

We describe the sample selection, observations, and data reduction in Sect. 2. In Sect. 3 we conduct a photometric analysis of our H$\alpha$ narrowband imaging, including deriving star formation rates (SFRs), eDIG scale heights ($h_{z\text{eDIG}}$), and H$\alpha$ radial profiles. We compare these results to a similar analysis of archival Galaxy Evolution Explorer (GALEX) far-ultraviolet (FUV) data in Sect. 4. In Sect. 5 we perform a kinematic analysis of our MUSE data, measuring ionized gas velocity lags and investigating the radial variations in the lag. We investigate the prevalent ionization mechanisms in Sect. 6 with emission-line diagnostics. The individual galaxies are discussed in Sect. 7. In Sect. 8 we discuss our results in the context of the ionization source and origin of the eDIG. The results of this work are summarized in Sect. 9.

\section{Observations and data reduction}

\subsection{The sample}
\begin{figure*}[ht]
  \centering
  \includegraphics[width=\textwidth]{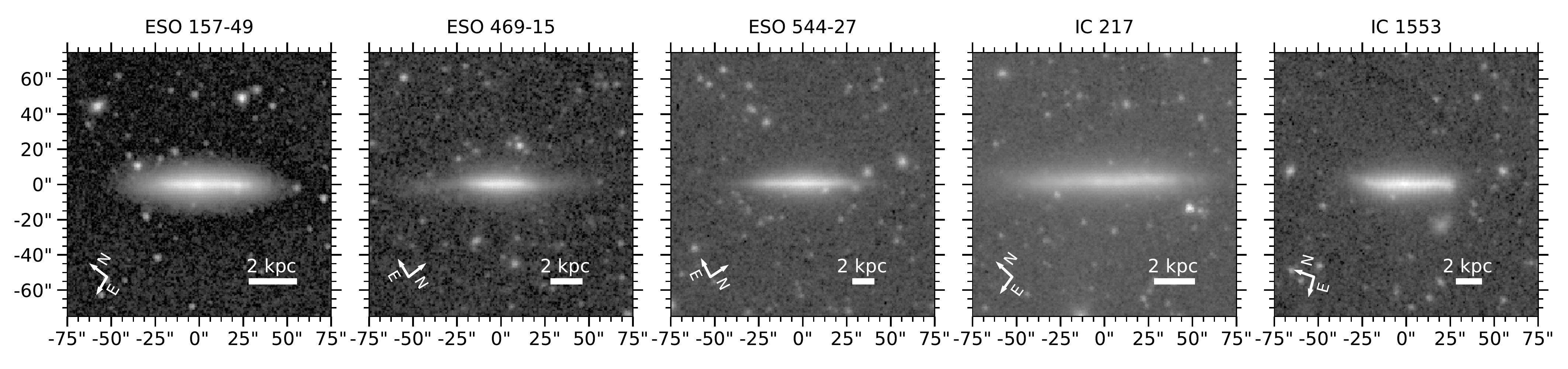}
  \caption{S$^4$G \citep{sheth2010s4g} 3.6 micron images of our sample galaxies. Directions on the sky and 2 kpc scale bars are shown on the images.}
  \label{fig:s4g}
\end{figure*}

Our sample is derived from that of \cite{comeron2019thick}, who observed eight edge-on galaxies with MUSE at the Very Large Telescope (VLT) of the European Southern Observatory (ESO), at Paranal Observatory, Chile. The galaxies were selected from the S$^4$G-selected 70 edge-on disk sample of \cite{comeron2012breaks} chosen to be observable from the Southern Hemisphere and to have an $r_{25} < 60\arcsec$ (as given in the HyperLEDA\footnote{\url{http://leda.univ-lyon1.fr/}} database; \citealt{makarov2014hyperleda}), so that the region between the center of the target and its edge could be covered in a single MUSE pointing. The quality of the data led to the serendipitous discovery of eDIG in all of the galaxies. In this paper we study the sample of five galaxies for which we were able to obtain deep narrowband imaging in addition to the MUSE data: \object{ESO~157-49}, \object{ESO~469-15}, \object{ESO~544-27}, \object{IC~217}, and \object{IC~1553}. Fundamental properties of the galaxies can be found in Table \ref{tab:stats} and S$^4$G images of the galaxies are shown in Fig. \ref{fig:s4g}. We plan to obtain narrowband imaging for two additional galaxies (PGC~28308 and PGC~30591), which will be included in future work.

\begin{table*}
  \caption{Basic properties of our galaxy sample.}
  \label{tab:stats}
  \centering
  \begin{tabular}{lccccccccccc}
    \hline\hline
    ID & PA & $h_{R}$ & $h_{z\text{t}}$ & $h_{z\text{T}}$ & \emph{d} & log$(M_*/M_{\odot})$ & $\nu_{\text{c}}$ & Telescope & \multicolumn{3}{c}{Exposure Time} \\
    & & & & & & &  & & H$\alpha$ & \emph{r} & FUV \\

    & ($^{\circ}$) & (kpc) & (kpc) & (kpc) & (Mpc) & & (km s$^{-1}$) & & (s) & (s) & (s) \\ 
    \hline
    ESO~157-49 & 30.4 & 0.97 & 0.10 & 0.45 & 17.3 & 9.81 & 107 & NTT & 27600 & 3390 & 1082 \\
    ESO~469-15 & 149.2 & 0.97 & 0.14 & 0.67 & 28.3 & 9.58 & 83 & NTT & 11050 & 780 & 1839\\
    ESO~544-27 & 153.3 & 1.81 & 0.16 & 0.67 & 45.9 & 9.84 & 129 & NOT & 43200 & 1080 & 565 \\
    IC~217 & 35.7 & 1.98 & 0.15 & 0.54 & 21.1 & 9.75 & 115 & NTT & 16200 & 1800 & 186 \\
    IC~1553 & 15.0 & 1.36 & 0.16 & 0.48 & 36.5 & 9.90 & 142 & NTT & 21600 & 2460 & 594 \\
    \hline
  \end{tabular}
  \tablefoot{Position angle (PA) values from \cite{salo2015s4g}. Scale lengths ($h_{R}$) and circular velocities ($\nu_{\text{c}}$) from \cite{comeron2019thick}. Thin disk scale height ($h_{z\text{t}}$) and thick disk scale height ($h_{z\text{T}}$) from \cite{comeron2018thick}. Distance estimates (\emph{d}; \citealt{tully2008dist,tully2016dist}) are the same as used in \cite{comeron2018thick, comeron2019thick}. Stellar masses ($M_*$) from \cite{munoz-mateos2015masses}. Also given are the telescope, the H$\alpha$ and the \emph{r}-band exposure times obtained for this work, and the FUV exposure times of the archival GALEX data.}
\end{table*}

\subsection{MUSE observations}
Our MUSE observations consist of four 2624 s on-target exposures for each galaxy except IC~217, for which three such exposures were taken. The exposures were centered on the same position for each galaxy, with 90\degree~rotations between exposures, totaling nearly three hours of exposure time for the covered areas (except for IC~217, which had two hours and ten minutes).

The data were reduced with the MUSE pipeline \citep{weilbacher2012pipeline} within the Reflex environment \citep{freudling2013reflex} and the combined datacubes were cleaned of sky residuals with the Zurich Atmosphere Purge (ZAP; \citealt{soto2016zap}). The datacubes were tessellated with the Voronoi binning code by \cite{cappellari2003voronoi} to produce bins with S/N $\sim$ 50 in H$\alpha$. Both stellar and emission-line velocities and velocity dispersions, as well as emission-line intensities were obtained with the Python version of the penalized pixel-fitting code (pPXF; \citealt{cappellari2004ppxf}). Further details of the MUSE observations, data reduction, Voronoi binning, and the pPXF processing can be found in \cite{comeron2019thick}.

\subsection{Narrowband H$\alpha$ imaging}
MUSE has a FoV of 1\arcmin $\times$ 1\arcmin. This is small compared to the angular size of the galaxies in our sample and, as such, additional observations were needed to gain critical information about the global spatial distribution of eDIG. We obtained deep narrowband H$\alpha$ imaging with the ESO Faint Object Spectrograph and Camera (EFOSC2) at the 3.58 m NTT at ESO's La Silla Observatory in Chile, and the Alhambra Faint Object Spectrograph and Camera (ALFOSC) at the 2.5 m NOT at the Observatorio del Roque de los Muchachos in Spain. EFOSC2 has a FoV of 4\farcm1 $\times$ 4\farcm1 and ALFOSC has a FoV of 6\farcm4 $\times$ 6\farcm4, although due to considerable vignetting caused by the filters, the effective FoV of ALFOSC is 4\farcm45 $\times$ 4\farcm45. These FoVs are large enough to easily cover our galaxies. We used EFOSC2's standard 2 $\times$ 2 binning readout mode, resulting in a pixel scale of 0\farcs24/px. No readout binning was used for ALFOSC, resulting in a pixel scale of 0\farcs2138/px.

We observed ESO~157-49, ESO~469-15, IC~217, and IC~1553 with the NTT over three nights in December 2018. As each galaxy had different observation windows, our total exposure times per galaxy ranged from three to eight hours in our on-band H$\alpha$ filters (filter number \#438 and \#761 depending on the galaxy recession velocity), with between 13 to 57 minutes in Gunn \emph{r} (filter \#786) to subtract the stellar continuum. We used filter \#761 for ESO~157-49, ESO~469-15 and IC~217, and filter \#438 for IC1553, with central wavelengths of 659 nm and 663 nm and full widths at half maximum (FWHMs) of 3 nm and 7 nm, respectively.

 We observed ESO~544-27 with the NOT in September 2020. We obtained 12 hours of on-target narrowband H$\alpha$ exposure (filter \#49) and 36 minutes of on-target Gunn \emph{r} exposure (filter \#14). The H$\alpha$ filter \#49 has central wavelength of 661 nm and a FWHM of 5 nm. The exact on-target exposure times for all galaxies are presented in Table \ref{tab:stats}.

 During the NTT data reduction we found that the standard flat-fielding procedure using twilight and dome flats was insufficient to produce uniformly flat backgrounds to our combined images. To avoid these large-scale patterns imposed by the flat-fielding, we constructed night-sky flat-fields from the science images. Due to the galaxies covering a significant portion of the FoV of EFOSC2 and the large amount of masking required because of this, the resulting flats had a relatively poor pixel-to-pixel accuracy. To avoid this issue with the NOT, we also observed a series of off-target blank sky exposures flanking the target field in order to construct the night-sky flats. We observed six hours of night sky in H$\alpha$ and 18 minutes of night sky in the \emph{r} band.

 We began our data reduction with a standard bias subtraction. We constructed preliminary flat-fields using dome flats taken during the day preceding the observations for the NTT data and exposures of the twilight sky for the NOT data, and applied the flat-field correction to the bias-subtracted science images. Using these preliminary flat-fielded images, we then constructed night-sky flats from both the science and blank sky exposures (for NOT data), following the method described by \cite{mihos2017flats}. To do this, we first masked astrophysical objects and scattered light artifacts, binned the images, and modeled the sky in each frame using a second-order polynomial fit to all unmasked pixels, using the Astropy\footnote{\url{http://www.astropy.org}} \citep{astropy:2013, astropy:2018} modeling module's Levenburg-Marquart least squares fitting routine.  We removed the sky gradients from each image by dividing the images with a normalized sky model, then combined these images into a new flat. We then flattened the images using that flat, remade and subtracted the sky models, and masked and binned the images again. We inspected these new images to see if any masks needed adjustment, to catch any diffuse light we missed during our first attempt at masking. If so, we adjusted, or simply did the gradient removal again, and combined the de-skied, masked images into a new flat. We iterated this process until the night-sky flat converged. Next, we flattened the images with this night-sky flat, modeled the skies one more time in the same manner (second-order polynomial), removed those skies from the images, altered the masks to include only leakages, such as scattered light artifacts and satellite trails, and then registered the images in right ascension and declination using Image Reduction and Analysis Facility (IRAF\footnote{IRAF was distributed by the National Optical Astronomy Observatory, which was managed by the Association of Universities for Research in Astronomy (AURA) under cooperative agreement with the National Science Foundation.}) software's WREGISTER task. To account for systematic errors in the frame-to-frame world coordinate system (WCS) registration, we applied offsets using the IMSHIFT task, aligning the images using centroids of a few handpicked stars. This ensures that the on-band and off-band images are perfectly aligned, which is needed for making the continuum-subtracted images. Before the coaddition, we scaled all of the images to a common zero point using the galaxy's total flux within a large elliptical aperture. Finally, we combined the images to create preliminary coadds.
 
Possibly due to suboptimal observing conditions -- some occurred with the moon above the horizon or during twilight hours, to optimize our time allocation -- the night skies in each frame contained complex structure not easily modeled using polynomial fits. To attempt to improve our sky subtraction, we thus employed a novel sky subtraction method using these preliminary coadds (al Sayyad \& Lupton, private communication). This involved aligning, flux-scaling, and subtracting the initial coadd from individual science exposures, to remove astrophysical flux and isolate the night sky. Due to the differences in the point spread functions (PSFs) between the initial coadd and the individual exposures, stars, H{~\sc ii} regions, and other thin structures like dust lanes are reduced in size but do not perfectly subtract out. Hence, to remove these from our sky maps, we eroded our initial object masks five to ten times using a circular kernel. Additionally, the sky maps generated by coadd-subtraction have extremely high pixel-to-pixel noise, and hence we binned the masked maps with bin sizes on the order of the target galaxy width, interpolated flux across any remaining masked pixels, and applied a Gaussian smoothing using a kernel with a FWHM of half the bin size in px. This results in a lower resolution but far less noisy map of the night sky in each frame, which we then subtracted as the individual frame skies. The resulting image backgrounds are significantly flatter than those produced using the polynomial fit skies. A new flat was constructed from the images again using the new sky models and divided out from the science images. This was again iterated until the new flat converges, and we created a final generation coadd using this final flat-field and these coadd-subtraction sky models. Because this sky subtraction method is novel, we will explore it in greater detail in a separate paper, including demonstrations of low-surface-brightness flux preservation using model galaxy injections. Preliminary tests show that it preserver flux at all surface brightnesses to within $\sim 5 \%$.

 To construct the continuum-subtracted H$\alpha$ images, the \emph{r}-band image must be scaled to be proportional to the stellar continuum emission in the H$\alpha$ filter image. We estimated this scale factor using two methods. First, we plotted the intensity of each pixel in the H$\alpha$ image against its intensity in the \emph{r}-band. Pixels with no line-emission have H$\alpha$ flux that is linearly proportional to the \emph{r}-band flux and appear as straight line in the plot. Pixels with H$\alpha$ emission have an excess flux over the stellar continuum, and so appear systematically above this line. The slope of the line gives the scale factor to scale the \emph{r} image to be subtracted from the H$\alpha$ image \citep{knapen2005diff,knapen2006diff}. Second, we compared the luminosities of the foreground stars in both filters and, under the assumption that those stars have no H$\alpha$ emission or absorption, took that as a secondary estimate of the scale factor for the \emph{r} image. These two methods give a range of values for the scale factor, from which we chose the final scale factor by eye so that over-subtraction and residual foreground starlight are minimized.

 We flux-calibrated the continuum-subtracted H$\alpha$ images by using the MUSE data. To do this we constructed mock continuum-subtracted narrowband H$\alpha$ images from the MUSE cubes by stacking a 60 Å band covering the H$\alpha$ and [N~{\sc ii}] lines and subtracting from it a nearby 60 Å band covering no emission lines. We then calibrated the flux so that the counts in several $\sim$ 1\arcsec-- 4\arcsec in radius circular apertures centered on H~{\sc ii} regions along the disk in the continuum-subtracted H$\alpha$ images are equal to those in the MUSE mocks. Both our MUSE and our narrowband observations reach an unbinned 1$\sigma$ background sensitivity of $\sim 10^{-17}$ erg s$^{-1}$ cm$^{-2}$ arcsec$^{-2}$ in H$\alpha$.

 \subsection{Archival GALEX data}
 To gain information about the age of the ionizing stellar population as well as the distribution of O and B stars, we retrieved archival GALEX FUV imaging from the Mikulski Archive for Space Telescopes (MAST). The GALEX FUV channel covers wavelengths from 1350 to 1750 Å, and the detector has a pixel scale of 1\farcs5/px and an effective angular resolution of ~6\arcsec. We use observations from the GALEX All-Sky Imaging Survey, Medium Imaging Survey, and the AKARI Deep Field South guest investigator program \citep{morrissey2007galex, bianchi2009galex}. The total exposure times per galaxy vary greatly due to the different depths of the used surveys, ranging from 186 to 1839 s. The exact exposure times are presented in Table \ref{tab:stats}. The GALEX data described here may be obtained from the MAST archive at \url{https://dx.doi.org/10.17909/t9-64rs-h623}.

\section{H$\alpha$ photometry}

\subsection{H$\alpha$ luminosity and star formation rates}
To investigate the connection between the eDIG ionization and SF, we must first obtain the SFR in our sample galaxies. To do this, we measured the integrated H$\alpha$ luminosity and the SFR of our galaxies from our narrowband imaging data. We identified sources that appear in the \emph{r}-band images, but not in the continuum-subtracted H$\alpha$ images; and sources that appear in the continuum-subtracted H$\alpha$ images, but show spectra inconsistent with H$\alpha$ emission of appropriate redshift in the MUSE cubes; as background or foreground sources and masked them out from the continuum-subtracted H$\alpha$ images. We also masked everything beyond where the galaxies blend into the background in the \emph{r} band images (pixel signal-to-noise drops down to the background value). We then measured the total H$\alpha$ flux from the masked images.

To correct for dust extinction, we measured the Balmer decrement using the integrated H$\alpha$ and H$\beta$ emission from the MUSE cubes and calculated the H$\alpha$ extinction adopting a Calzetti extinction law \citep{calzetti2000extinction} with $R_{V}=4.5$ \citep{fischera2005rv}. The MUSE pointings cover each galaxy only partially, the pointings being centered on one side of the galaxy if the whole disk does not fit into a single MUSE pointing. While this does increase the uncertainty of our extinction estimates, it does not do so significantly, if we believe the reasonable assumption that the dust distribution and extinction has some degree of azimuthal symmetry. We calculated the extinction-corrected H$\alpha$ luminosity using distances from \cite{tully2008dist,tully2016dist}. The luminosities are presented in Table \ref{tab:photo}.

\begin{table*}
  \caption{Photometry results.}
  \label{tab:photo}
  \centering
  \begin{tabular}{lccccccccccc}
    \hline\hline
    ID & log$(L_{\text{H$\alpha$}}/L_{\odot})$ & SFR & log(sSFR/yr) & $h_{z\text{H {\sc ii}}}$ & $h_{z\text{DIG}}$ & $h_{z\text{eDIG}}$ & $(\chi^2 / \nu)_2$ & BIC$_2$ & $h_{z1}$& $(\chi^2 / \nu)_1$ & BIC$_1$ \\
     & & ($M_{\odot}$/yr) &  & (kpc) & (kpc) & (kpc) & & & (kpc) & & \\ 
    \hline
    ESO~157-49 & 6.97 & 0.19 & -10.5 & 0.20 & 0.23 & 0.68 & 2.24 & 854 & 0.31 & 4.29 & 1612 \\
    ESO~469-15 & 6.98 & 0.20 & -10.3 & 0.13 & 0.12 & 1.31 & 16.06 & 3732 & 0.14 & 21.42 & 4985 \\
    ESO~544-27 & 6.81 & 0.13 & -10.7 & 0.22 & 0.20 & 0.59 & 2.74 & 675 & 0.35 & 4.83 & 1170 \\
    IC~217 & 7.09 & 0.25 & -10.3 & 0.22 & 0.22 & 1.16 & 13.07 & 3996 & 0.24 & 16.48 & 5043 \\
    IC~1553 & 7.67 & 0.96 & -9.92 & 0.25 & 0.24 & 1.39 & 4.14 & 1123 & 0.30 & 22.04 & 5902 \\
    \hline
  \end{tabular}
  \tablefoot{The H$\alpha$ luminosities ($L_{\text{H$\alpha$}}$) are extinction-corrected using the Balmer decrement. The star forming rates (SFR) are calculated using the \cite{hao2011sfr} and the \cite{murphy2011sfr} calibrations. The sSFRs are calculated using stellar masses from \cite{munoz-mateos2015masses}. The scale heights are given for the H~{\sc ii} regions ($h_{z\text{H {\sc ii}}}$), the quiescent ($h_{z\text{DIG}}$) and disturbed extraplanar ($h_{z\text{eDIG}}$) DIG components of the double-exponential fit, and the single-exponential DIG fit ($h_{z1}$). The $\chi^2$ and BIC values demonstrate the goodness of fit for our double-exponential (2) and single-exponential (1) fits.}
\end{table*}

\begin{figure}
  \centering
  \includegraphics[width=0.5\textwidth]{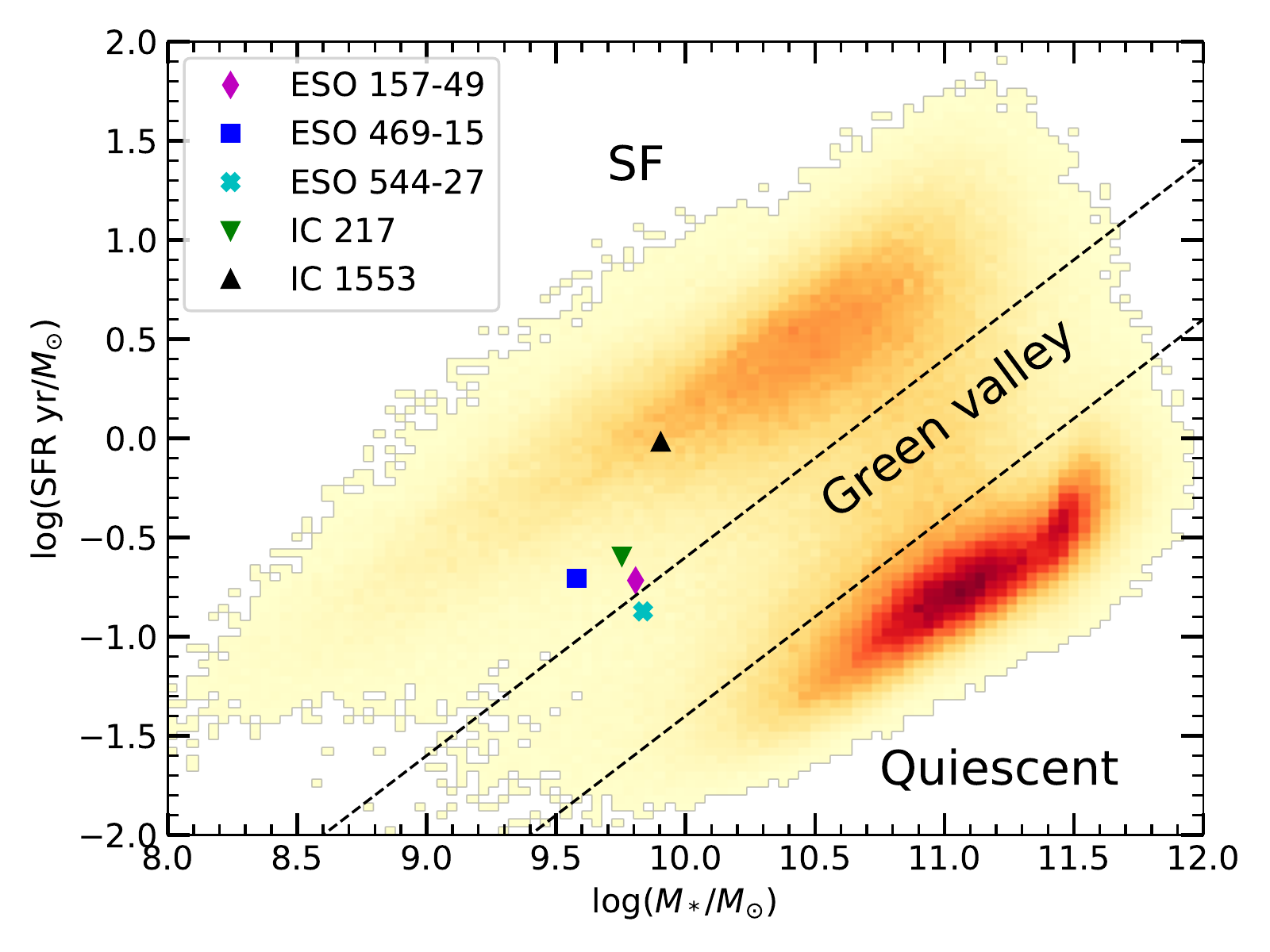}
  \caption{Stellar mass against SFR relation for our sample. The heatmap represents Sloan Digital Sky Survey (SDSS) galaxies with $M_*$ from \cite{kauffmann2003smass} and \cite{salim2007mass} and the SFR from \cite{brinchmann2004sfr}. The dashed lines correspond to constant sSFRs of $10^{-10.6}$ and $10^{-11.4}$ yr$^{-1}$. These lines separate star-forming galaxies, the green valley, and quiescent galaxies \citep{hsieh2017manga}.}
  \label{fig:ms}
\end{figure}

We calculated the SFR from the extinction-corrected H$\alpha$ luminosities using the \cite{hao2011sfr} and the \cite{murphy2011sfr} calibrations. We find SFR values ranging from 0.13 to 0.96 $M_{\odot}\text{yr}^{-1}$. We also calculated the specific SFR (sSFR = SFR$M_*^{-1}$) using 3.6 $\mu$m stellar masses from \cite{munoz-mateos2015masses}. The SFR and sSFR are presented in Table \ref{tab:photo} and the SFR--$M_*$ relation for our sample is shown in Fig.~\ref{fig:ms}. Of our galaxies, only IC~1553 is firmly in the star-forming main sequence \citep{noeske2007ms}. ESO~157-49, ESO~469-15, and IC~217 are also marginally star-forming when star-forming galaxies are defined as those with sSFR > $10^{-10.6}$ yr$^{-1}$ \citep{hsieh2017manga}, while ESO~544-27 falls into the green valley.

\subsection{Vertical profiles and scale heights}
\label{sec:vprof}
To characterize eDIG and its spatial distribution, we obtained vertical H$\alpha$ profiles from the narrowband data. We estimated the alignment of the disk planes using the position angles from \cite{salo2015s4g}. We also masked ESO~469-15's extraplanar H{~\sc ii} regions so they do not influence the eDIG measurements.

To obtain the vertical profiles of the ionized gas, we averaged the H$\alpha$ intensity at each height over the plane of the galaxy in one-pixel-wide bins. We radially restricted the measurement to areas with clear visually identifiable midplane H~{\sc ii} regions. We then modeled the vertical profile with two exponential disks. We also attempted to fit the vertical profile at different radii in one scale length bins but find that our data are not sensitive enough for well-constrained two-component fits when not integrating over the whole disk.

The shape of the surface brightness profiles is also affected by the PSF of the data. To obtain the PSF we adopted a method similar to \cite{infante-sainz2020psf}, using stars with different brightnesses to measure different parts of the PSF. For the NTT data, we modeled the PSF by combining an averaged core PSF of several bright stars in the data with mid-range PSF obtained from the brightest star in the data. We do not have bright enough stars in our data to measure the arcminute-scale extended wings of the PSF, but because it is primarily these faint wings that bias the light profiles of galaxies at low surface brightness \citep{rudick2009psf, sandin2014psf, sandin2015psf}, we instead extrapolated the measured PSF down below the image noise limits using mock wings following a $r^{-2}$ power-law. A power-law $r^{-2}$ was chosen as it is a common slope in the PSF wings of ground based observations \citep{king1971psf, slater2009psf, sandin2014psf}. We experimented with other values for the wing slopes, ranging from $r^{-1.5}$ to $r^{-2.5}$, and found the effect to be negligible to our results. The averaged core PSF extended out to 3\farcs6, the middle section from the single bright star to 16\farcs9, and the power law wings out to 150\arcsec.

\begin{figure}
  \centering
  \includegraphics[width=0.5\textwidth]{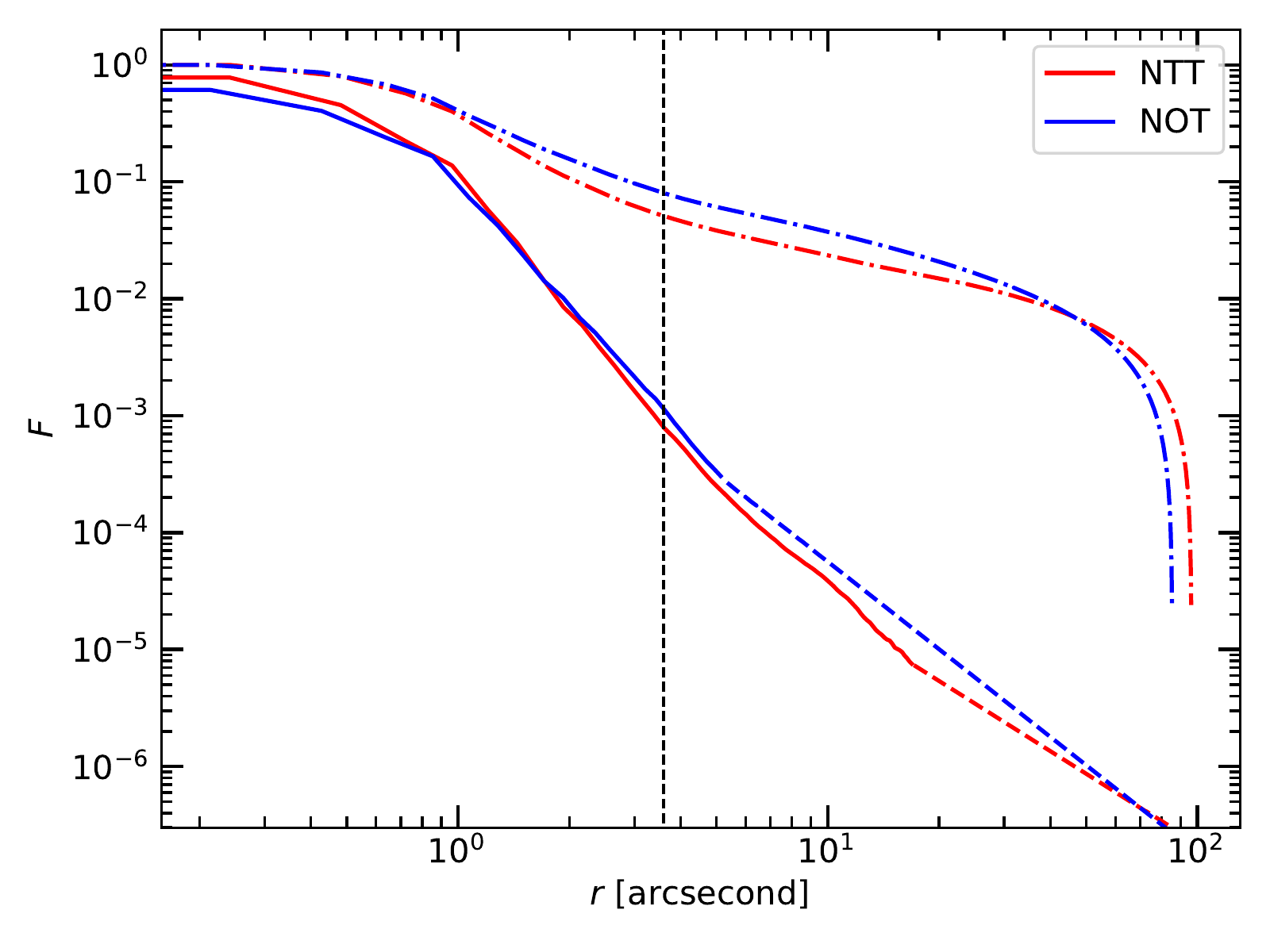}
  \caption{PSF radial profiles measured and extrapolated from our NTT (red lines) and NOT (blue lines) data. The dashed lines show the extrapolated power-law wings. The dot-dashed lines indicate the fractional cumulative flux outside a given distance from the center. The black vertical dashed line corresponds to the separation between the averaged core PSF and the bright star middle section at 3\farcs6 for the NTT PSF. The NTT PSF is normalized to its maximum value, and the NOT PSF is normalized so that it has a total flux equal to the NTT PSF.}
  \label{fig:psf}
\end{figure}

We attempted to model the NOT PSF in a similar manner, but our observations with this telescope lacked sufficiently bright stars to create this three-stage PSF. Instead we made the PSF from two stages of the averaged core within 5\farcs3 and a mock power law wings beyond that. The wings are clearly steeper for the NOT data so we used power-law of $r^{-2.5}$. Radial profiles of the PSFs are shown in Fig. \ref{fig:psf}.

We created 2D images of our double-exponential model, then convolved them with our 2D PSFs. We then extracted the vertical profiles from these convolved model images and fit these to the measured vertical profiles, leaving the scale heights and central fluxes of the two components, as well as the midplane position ($z=0$) as free parameters. We interpret the scale height of the thinner component as the vertical extent of the in-plane H{~\sc ii} regions ($h_{z\text{H {\sc ii}}}$ in Table \ref{tab:photo}). Then we fit the profile again but masked the previously obtained H{~\sc ii} region dominated thin disk region $|z| < h_{z\text{H {\sc ii}}}$ and constrain the midplane to the same $z$ as obtained from the first fit. For the second fit we adopt the interpretation of \cite{miller2003twodisk}, where the thinner component of the remaining H$\alpha$ emission is the brighter quiescent phase of DIG and the thicker component is the fainter disturbed extraplanar phase of DIG, or eDIG. We used Poisson weighting in the fits. The scale heights obtained from these fits are presented in Table \ref{tab:photo} ($h_{z\text{DIG}}$ and $h_{z\text{eDIG}}$) and the fits and profiles themselves are shown in Fig. \ref{fig:prof}. For the quiescent component we obtain scale height values ranging from 0.12 to 0.24 kpc and for the disturbed eDIG from 0.59 to 1.39 kpc.

\begin{figure*}[ht]
  \centering
  \includegraphics[width=\textwidth]{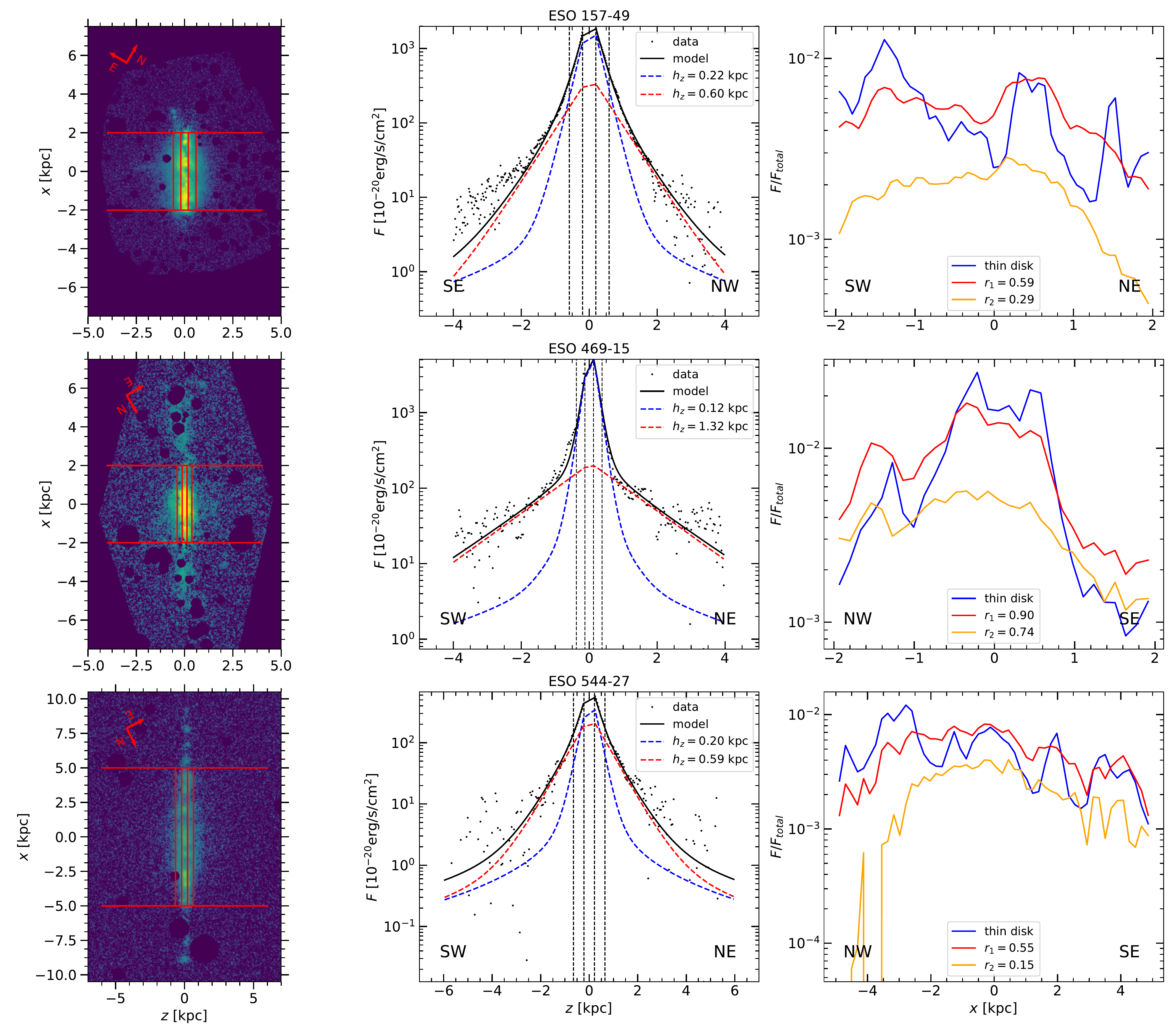}
  \caption{Masked narrowband H$\alpha$ images (left) and vertical (center) and radial (right) H$\alpha$ profiles for our sample. Black dots in the center column show measured fluxes, and the black curves show our two-component exponential fits. The dashed blue curve in the center column is the thinner component of the fit, representing the quiescent phase of the DIG. The dashed red curve in the center column is the thicker component of the fit, representing the disturbed eDIG. The fit scale heights are shown for both components in the legends of the center column. The vertical dashed black lines in the center column are the inner limits of lenient (inner) and stringent (outer) eDIG definitions (see text). These are also shown in the H$\alpha$ images of the left column via the vertical solid red lines. The horizontal solid red lines in the left column show the radial extent of the thin disk and eDIG definitions. In the right column the solid blue lines represents the thin disk radial H$\alpha$ profiles, the solid red lines represents the eDIG radial H$\alpha$ profiles using the lenient eDIG definition (eDIG$_1$), and the solid orange lines represent the eDIG radial H$\alpha$ profiles using the stringent eDIG definition (eDIG$_2$). The thin disk and lenient eDIG radial profiles are normalized to unity flux, and the stringent eDIG radial profiles are normalized to the flux of the lenient eDIG flux to highlight the similarities between the profiles. The disk--eDIG Pearson \emph{r} values are shown in the legends of the right column. Directions on the sky are indicated in the images and the plots.}
  \label{fig:prof}
\end{figure*}

\begin{figure*}[ht]
  \centering
  \includegraphics[width=\textwidth]{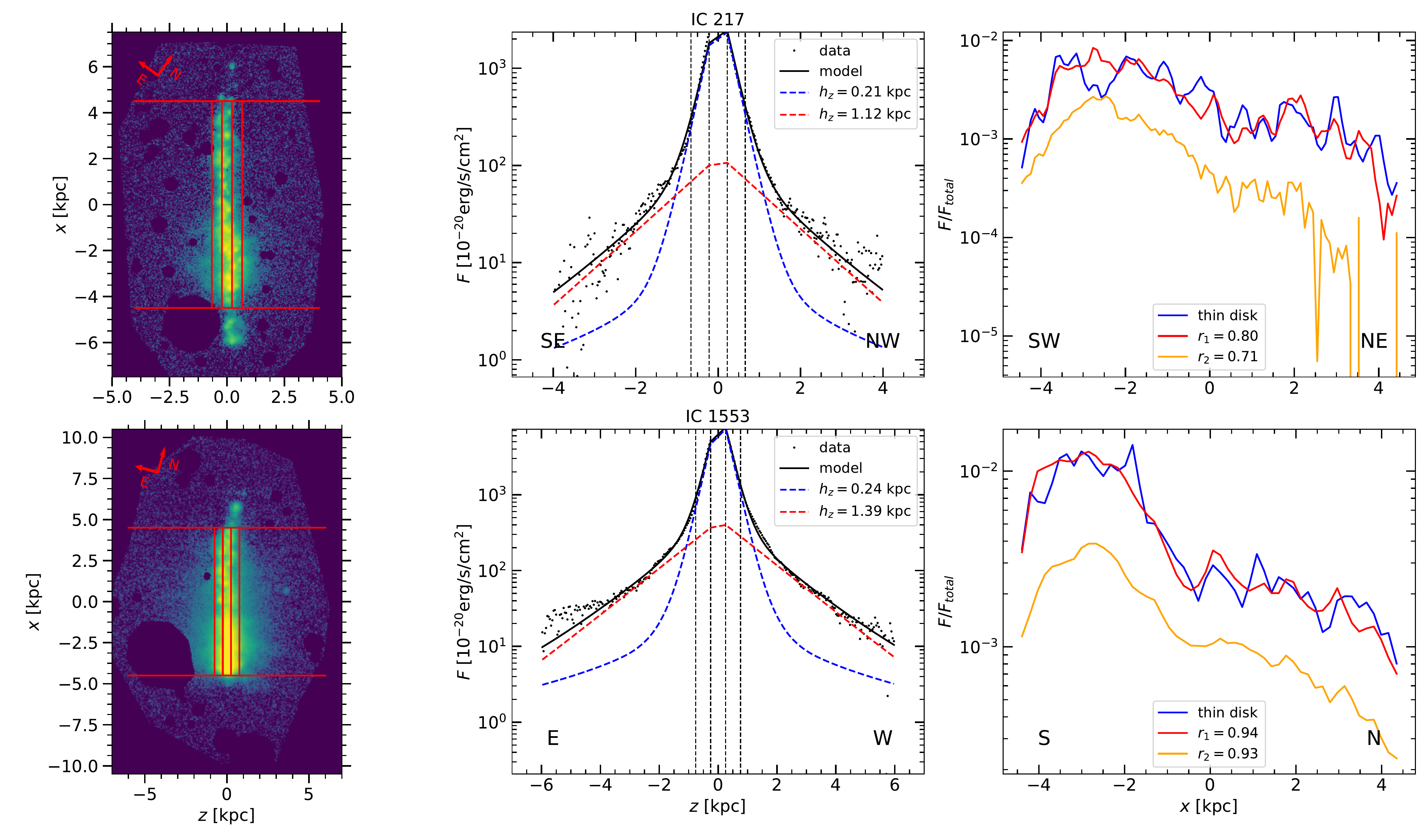}
  \addtocounter{figure}{-1}
  \caption{Continued.}
\end{figure*}

\begin{figure}
  \centering
  \includegraphics[width=0.5\textwidth]{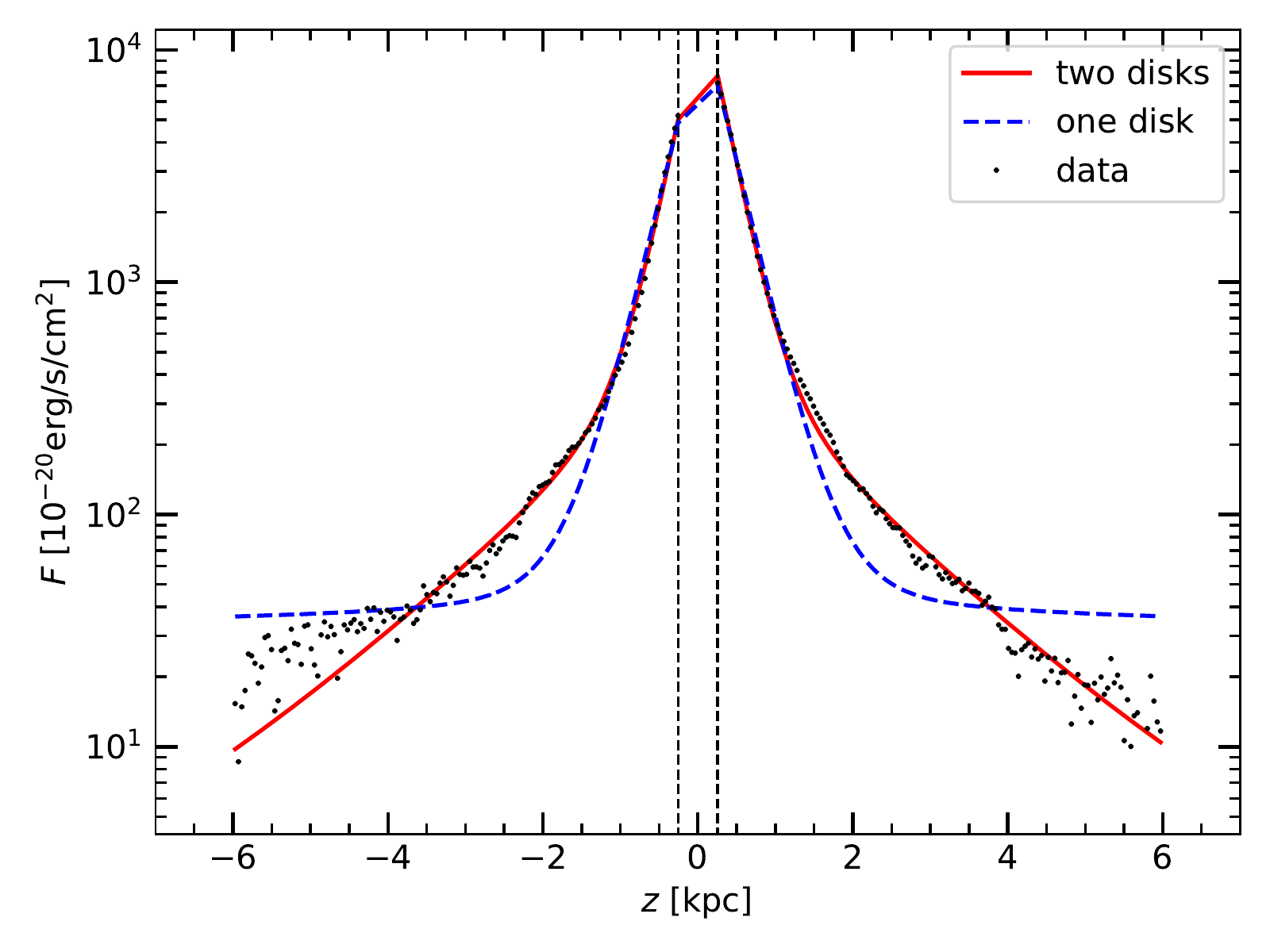}
  \caption{Comparison of vertical H$\alpha$ profile fits of IC~1553 using our double-exponential model (solid red line) and single-exponential model with a constant background (dashed blue line). Black dots show the measured fluxes, and the black vertical dashed lines show the extent of the thin disk.}
  \label{fig:profcomp}
\end{figure}

Most previous eDIG scale height measurements in literature have relied on single-exponential fits due to insufficient sensitivity. To better compare our results to these previous works, we obtained single-exponential H$\alpha$ scale heights from our data in the same manner we obtained the double-exponential scale heights, except we modeled the vertical profiles with a single-exponential disk and a constant background instead of two exponential disks. The single-exponential fits give scale heights ranging from 0.14 to 0.35 kpc ($h_{z1}$ in Table \ref{tab:photo}). Our scale height measurements were performed on sky-corrected images where the background should be near zero, but the single-exponential fits require positive backgrounds with surface brightness on the order of our sensitivity limit. This may be caused by the emission from the disturbed eDIG affecting the backgrounds, suggesting that a single-exponential fits are insufficient in modeling the H$\alpha$ vertical profiles of our sample galaxies. We also experimented with the \cite{kruit1988prof} three-parameter vertical density profile\footnote{$I(z) = 2^{-2/n} I_0 \text{sech}^{2/n}(nz/2z_0)$}, but found that the fits tended toward the limiting case of an exponential disk ($n = \infty$). Fig. \ref{fig:profcomp} illustrates how single-component fits fail to model simultaneously both the steep quiescent DIG near the midplane and the extended disturbed eDIG at large heights.

To test the goodness of fit of our models we calculated the $\chi^2$ and Bayesian information criterion \citep[BIC; see][]{schwarz1978bic} values for our double-exponential and single-exponential fits. The $\chi^2$ is defined as
  \begin{equation}
    \chi^2 = \sum_z^{n_z} \frac{[I_{obs}(z) - I_{model}(z)]^2}{\sigma(z)^2}
  ,\end{equation}
where $I_{obs}$ is the observed intensity of the profile, $I_{model}$ is the intensity of the fitted model, $\sigma$ is the Poisson uncertainty of the measured profile calculated using background standard deviation and the effective gain of the combined image, and the sum is calculated over the vertical profile. If the data are well described by the model, $\chi^2$ will be approximately equal to the number of degrees of freedom $\nu = n_z - k$, where $n_z$ is the number of vertical bins in the profiles and $k$ is the number of free parameters in the models. The latter is four for our double-exponential model (central intensities and scale heights of the two disk) and three for our single-exponential model (central intensity and scale height of the disk and the intensity of the constant background). The BIC, formulated \citep[according to][]{kass1995bic} as
  \begin{equation}
    \text{BIC} = \chi^2 + k \ln n_z
  ,\end{equation}
is based on the $\chi^2$, but adds a second term that penalizes model complexity. This way, BIC is a good statistic when comparing models with different numbers of components, giving larger (worse) values to overfitted models. The normalized $\chi^2$ values ($\chi^2 / \nu$) and BIC values for our double-exponential and single-exponential fits are given in Table \ref{tab:photo}. The double-exponential fits have smaller (better) $\chi^2 / \nu$ and BIC values than the single-exponential fits for all our galaxies, indicating that the double-exponential fits represent our data more accurately. However, we find $\chi^2 / \nu$ values much greater than one for even our double-exponential fits, suggesting that neither double-exponential nor single-exponential disks describe the eDIG very well. This is not surprising considering the filamentary and asymmetric eDIG morphology present in our sample galaxies. While scale heights derived from exponential disk models are a useful tool to measure the vertical extent of eDIG, it is clear that simple exponential disks cannot describe eDIG fully.

\begin{figure}
  \centering
  \includegraphics[width=0.5\textwidth]{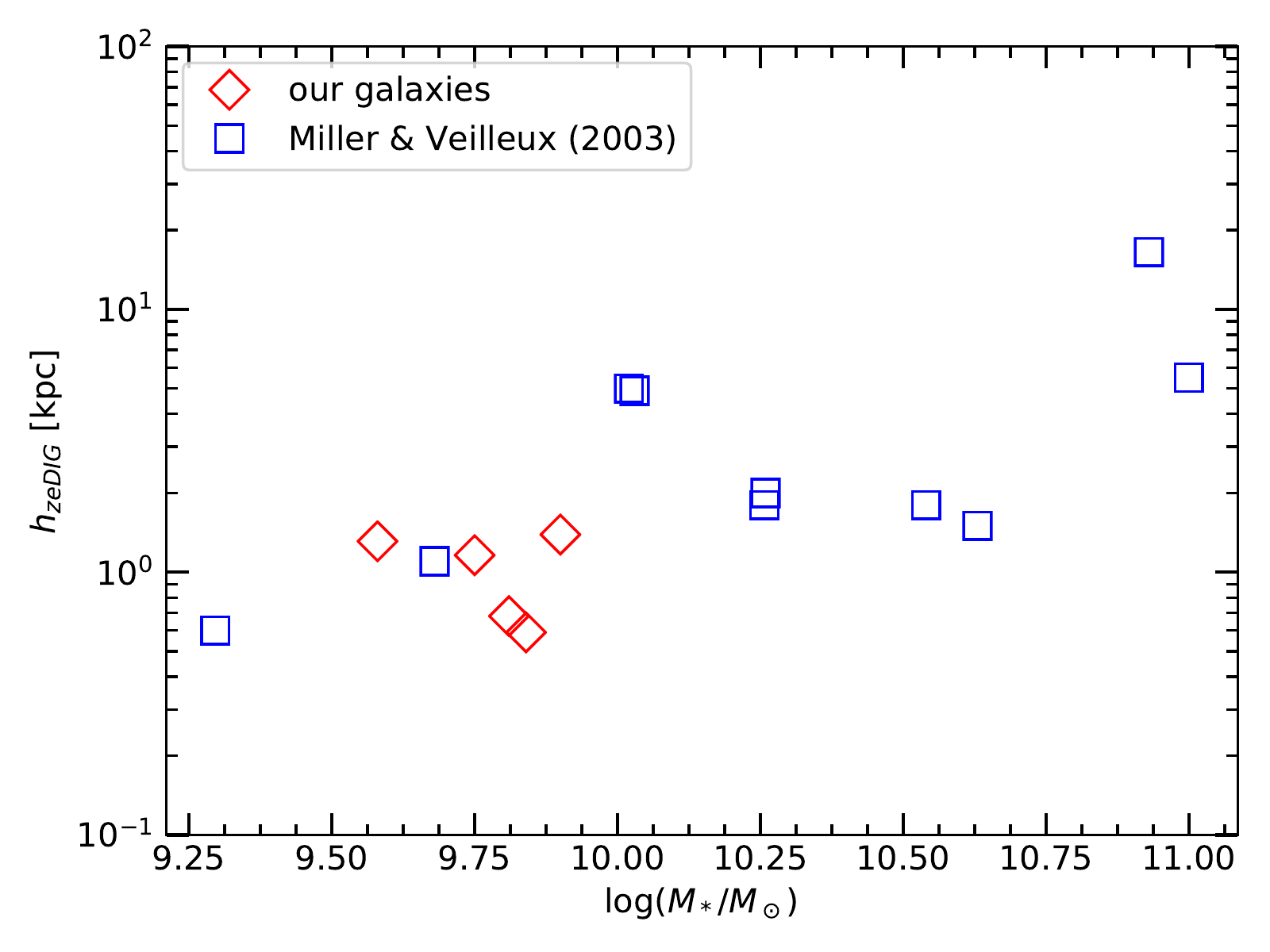}
  \caption{Stellar mass against eDIG scale height for our sample galaxies (red diamonds) and for a subset of \cite{miller2003twodisk} sample galaxies for which stellar masses were derived by \cite{munoz-mateos2015masses} (blue squares).}
  \label{fig:MV}
\end{figure}

\cite{miller2003twodisk} use double-exponential fits to the H$\alpha$ vertical profile and find a median scale height of $2.3 \pm 4.3$ kpc and a range of values from 0.4 to 17.9 kpc for the disturbed eDIG component over their sample of 16 galaxies. Fig. \ref{fig:MV} shows the scaling relation between $M_*$ and $h_{z\text{eDIG}}$ for our sample and for those galaxies of \cite{miller2003twodisk} for which stellar masses were derived by \cite{munoz-mateos2015masses}. \cite{levy2019edge} find a median scale height of $0.8^{+0.7}_{-0.4}$ kpc with single-exponential fit over their sample of 25 edge-on galaxies from the Calar Alto Legacy Integral Field Area Survey (CALIFA). They measure scale heights ranging from 0.3 to 2.9 kpc in galaxies with stellar masses ranging from 10$^9$ to 10$^{11}$ $M_{\odot}$. Furthermore, \cite{levy2019edge} compiled an extensive list of eDIG scale heights from various sources \citep{rand1997digspec1, wang1997dig, hoopes1999dig, collins2000dig, collins2001dig, miller2003twodisk, rosado2013eso379, bizyaev2017manga} and found a median eDIG scale height of $1.0 \pm 2.2$ kpc. While there is considerable scatter from galaxy to galaxy in the eDIG measurements in literature, our values -- using both double-exponential and single-exponential fits -- are still well within the large range of previous scale height measurements. 

\subsection{Radial profiles}
\label{sec:rad}
To complete our characterization of the eDIG with the spatial distribution along the radial axis, we measured the radial profiles of the ionized gas both in the thin disk and in the extraplanar regions. For the thin disk we averaged the H$\alpha$ intensity in 4 pixel wide radial bins over the vertical extent of the H~{\sc ii}-region-dominated thin disk, using the same thin disk region described in Sect. \ref{sec:vprof} ($|z|< h_{z\text{H {\sc ii}}}$). The width of the radial bins (4 pixels) was chosen to be on the order of the PSF FWHM, in order to minimize PSF effects along the radial axis. For the extraplanar gas we used two different eDIG area definitions to investigate the effects of vertical PSF contamination from the bright disk H~{\sc ii} regions. For a lenient eDIG definition (eDIG$_1$) we averaged the radial H$\alpha$ intensity over the heights above and below the thin disk up to the background mask ($|z|> h_{z\text{H {\sc ii}}}$). For a stringent eDIG definition (eDIG$_2$) we averaged the radial H$\alpha$ intensity from $|z| > 3 h_{z\text{H~{\sc ii}}}$ to the background mask, leaving a $2 h_{z\text{H~{\sc ii}}}$ buffer region between the thin disk and the stringent eDIG that is not used in either radial profile. The radial profiles are shown in Fig. \ref{fig:prof}.

\begin{figure*}[ht]
  \centering
  \includegraphics[width=\textwidth]{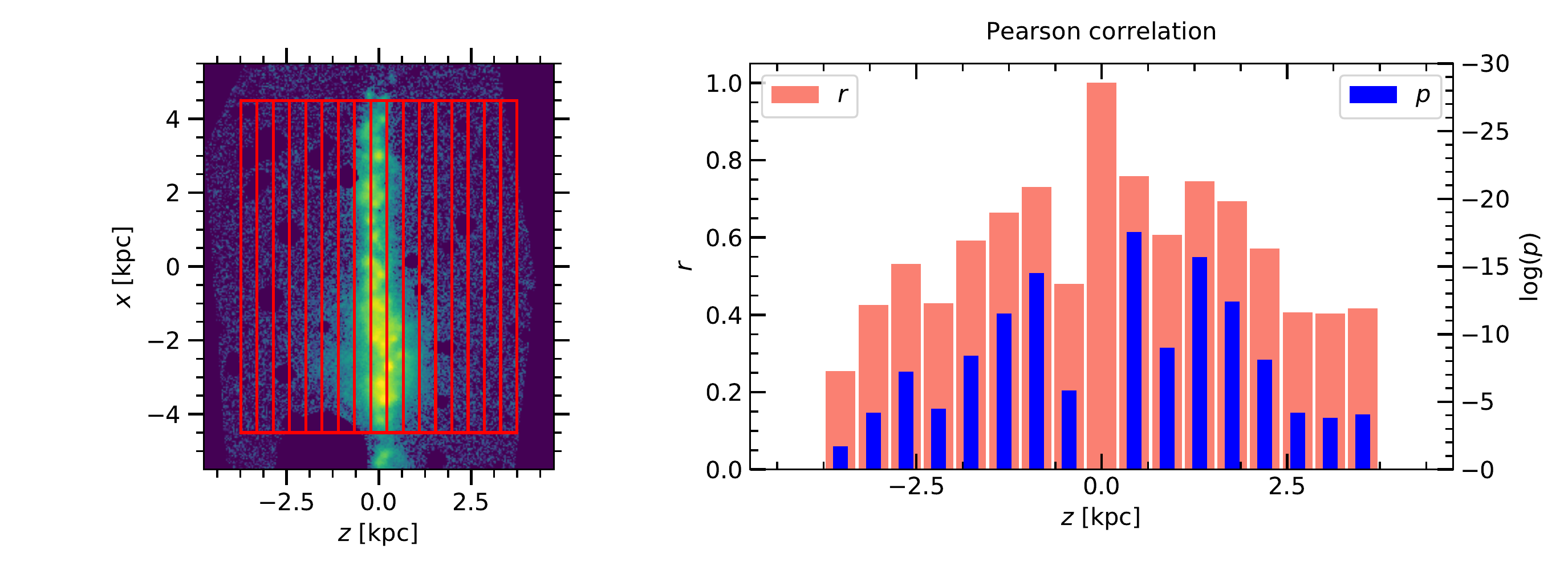}
  \caption{Disk--eDIG radial profile Pearson correlation at different heights for IC~217. The left panel shows the H$\alpha$ image of the galaxy. The right panel shows the correlation coefficient (\emph{r}; red bars) and the $p$-value ($p$; blue bars) at different heights (\emph{z}). The red rectangles in the left image correspond to the bins from which the eDIG radial profiles at different heights were measured.}
  \label{fig:pearson}
\end{figure*}

We then calculated the Pearson correlation coefficients between the thin disk radial profile and the two eDIG radial profiles with the intention of quantifying the spatial correlation between the disk H~{\sc ii} regions and eDIG. A similar approach using a single eDIG definition was done by \cite{miller2003twodisk}. The Pearson correlation coefficients between the H~{\sc ii}-region-dominated disk and the lenient and stringent eDIG (\emph{r} values) are given in Table \ref{tab:rad}. A correlation coefficient of 1 (-1) would signify a perfect (anti)correlation, while 0 would mean no correlation. Using the lenient eDIG definition all galaxies show some degree of correlation with the lowest \emph{r} value being 0.55 and the highest 0.94. Using the stringent eDIG definition does not greatly affect the correlation in the galaxies with the strongest correlations (ESO~469-15, IC~217, and IC~1553), but it does significantly weaken the correlation in the galaxies with the weakest correlations (ESO~157-49 and ESO~544-27), with a minimum of 0.15 in the case of ESO~544-27. This shows that while the correlation between the lenient eDIG and the thin disk may be strongly influenced by PSF in ESO~157-49 and ESO~544-27, this is clearly not the case for ESO~469-15, IC~217, and IC~1553. It should be noted that regardless of PSF contamination, the correlation coefficients between the eDIG and the thin disk radial profiles depend on the height at which the eDIG radial profile is measured, as illustrated by Fig. \ref{fig:pearson}. Also given in Table \ref{tab:rad} are the correlation \emph{p}-values, which roughly indicate the probability of an uncorrelated system producing data sets with at least as high Pearson $r$ as the data set in question. The highest \emph{p}-value for our correlations is $\sim$0.28 for the stringent eDIG of ESO~544-27, indicating a 28\% probability of the radial profiles being uncorrelated, while the other \emph{p}-values indicate probabilities of less than 5\%. There is clearly a significant positive correlation between the disk and eDIG H$\alpha$ emission, suggesting that leaky H~{\sc ii} regions are indeed an important source of ionization for the eDIG.

\begin{table}
  \caption{Radial profile correlations.}
  \label{tab:rad}
  \centering
  \begin{tabular}{lcccc}
    \hline\hline
    ID & $r_1$ & $r_2$ & log($p_1$) & log($p_2$) \\
    \hline
    ESO~157-49 & 0.59 & 0.29 & -5.00 & -1.34 \\
    ESO~469-15 & 0.90 & 0.74 & -10.8 & -5.49 \\
    ESO~544-27 & 0.55 & 0.15 & -4.54 & -0.56 \\
    IC~217 & 0.80 & 0.71 & -20.8 & -14.4 \\
    IC~1553 & 0.94 & 0.93 & -24.3 & -22.6 \\
    \hline
  \end{tabular}
  \tablefoot{The Pearson correlation coefficients ($r_1$, $r_2$) and associated p-values ($p_1$, $p_2$) demonstrate correlations between the radial profiles of the thin disk and our lenient (1) and the stringent (2) definitions of extraplanar H$\alpha$ emission (see Sect. \ref{sec:rad}).}
\end{table}

Dust obscuration is much more impactful to the disk radial H$\alpha$ profile compared to the eDIG profiles. To test whether dust obscuration is weakening our correlations, we derived raw and extinction corrected (using the Balmer decrement) H$\alpha$ maps from our MUSE cubes, and performed the radial profile correlation analysis described above on them. Due to the MUSE FoV being smaller than the radial extent of some of our sample galaxies, the MUSE cube derived profiles are not comparable to our narrowband imaging profiles. However, differences between the MUSE-derived extinction-corrected and raw H$\alpha$ maps should demonstrate the effects of dust obscuration on our analysis.  We do not find significantly stronger correlations on the extinction corrected H$\alpha$ maps compared to the raw H$\alpha$ maps for any of our galaxies. On the contrary, most correlations are slightly weakened on the extinction corrected images. This is most likely caused by the increased noise on the extinction corrected H$\alpha$ maps. Dust obscuration does not seem to have significant effect on our radial profile correlation analysis.

\subsection{Trends with SFR}
Now that we have both obtained the SFR and characterized the eDIG spatial distribution in our sample, we can investigate their connection. Correlations between the eDIG properties and the SFR are expected if the eDIG is a result of star-forming activity. If leaky H{~\sc ii} regions in the thin disk drive the eDIG ionization, one would expect a positive correlation between the sSFR and the eDIG scale height. If other mechanisms also play part in the ionization of eDIG, one might expect the disk and the eDIG H$\alpha$ emission to be more strongly spatially correlated in galaxies with higher sSFR.

\begin{figure}
  \centering
  \includegraphics[width=0.5\textwidth]{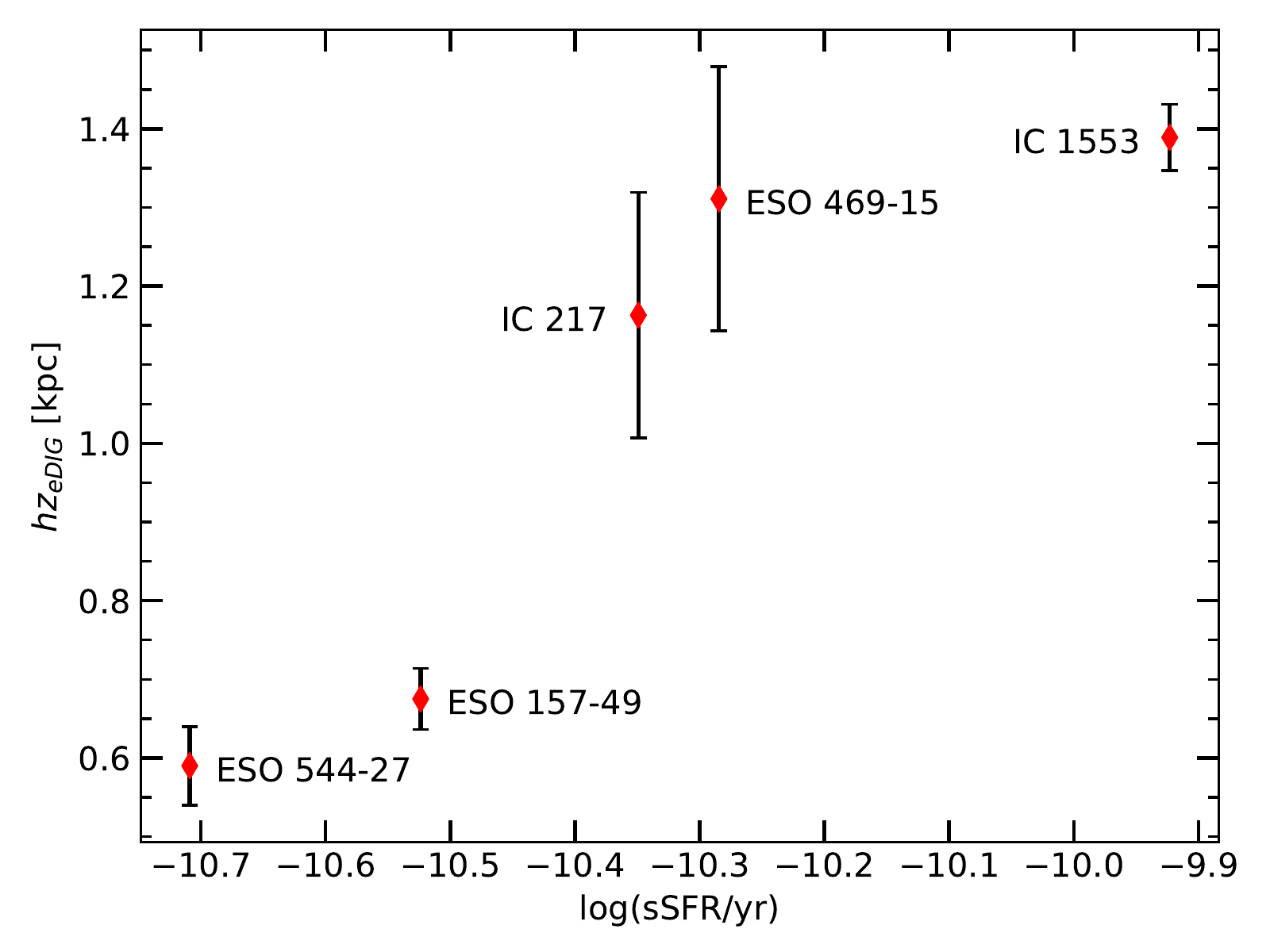}
  \caption{Specific SFR against eDIG scale height for our sample. One-standard-deviation error bars of the $h_{z\text{eDIG}}$ fit are shown.}
  \label{fig:edig}
\end{figure}

\begin{figure}
  \centering
  \includegraphics[width=0.5\textwidth]{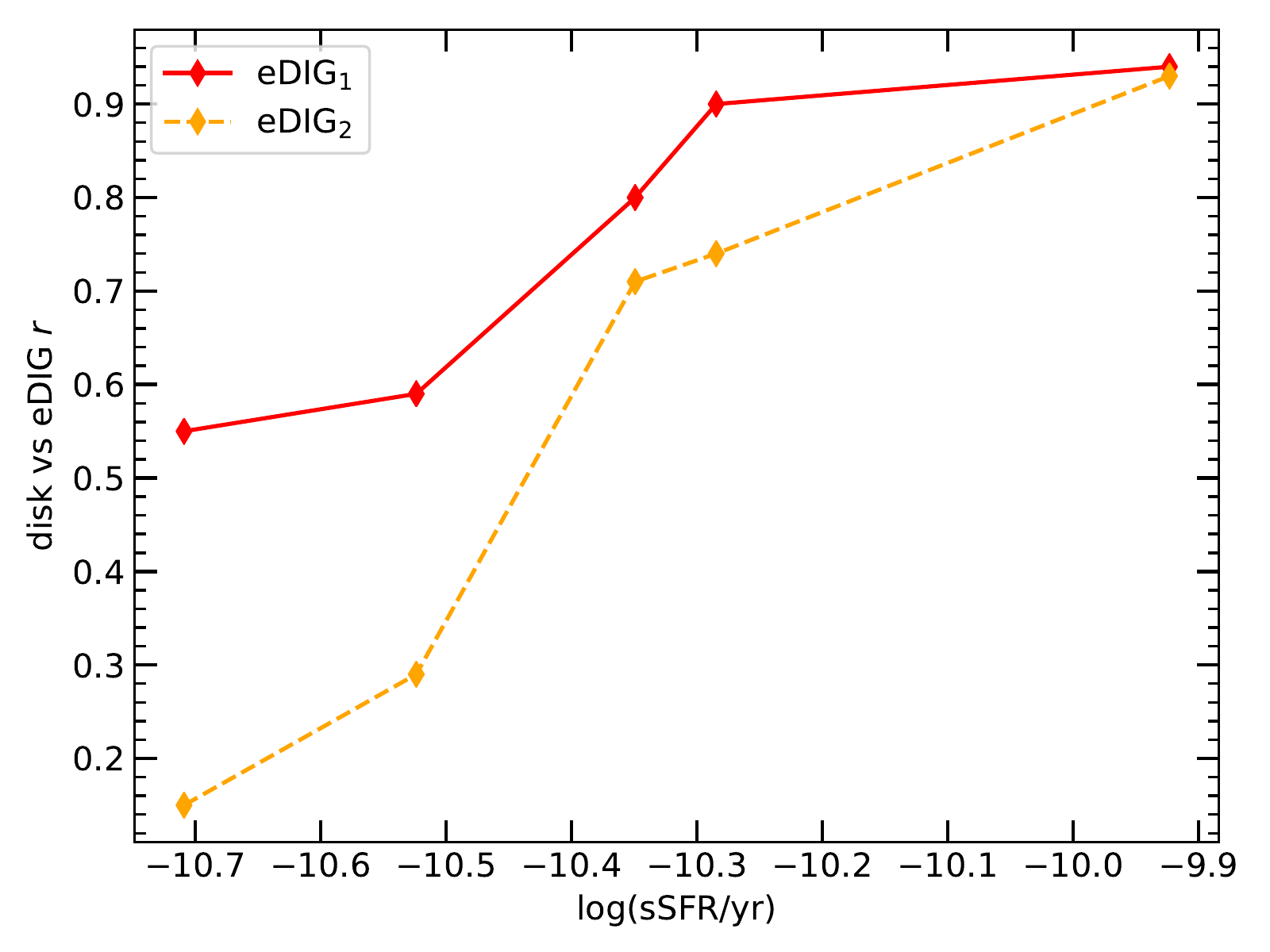}
  \caption{Specific SFR against disk--eDIG radial profile Pearson correlation for our sample. The solid red line is computed using our lenient eDIG definition (eDIG$_1$) and the dashed orange line using our stringent eDIG definition (eDIG$_2$).}
  \label{fig:pearson2}
\end{figure}

We compare the measured eDIG scale heights and disk--eDIG \emph{r} values to sSFR in Figs. \ref{fig:edig} and \ref{fig:pearson2}. There appears to be a clear positive correlation between the sSFR and the eDIG scale height, supporting a link between the disk SF and the eDIG ionization. There is also a positive correlation between sSFR and disk--eDIG \emph{r} values. For the disk--eDIG \emph{r} values, using the stringent eDIG area definition less affected by the PSF contamination results in an even stronger correlation with the sSFR. The positive correlation between sSFR and disk--eDIG \emph{r} values suggests that in galaxies with less SF, other mechanisms may contribute to the eDIG ionization significantly, and thus weaken the correlation between disk and eDIG H$\alpha$ emission. Given our small sample size ($N=5$), we admit that these correlations may be spurious. However, we find them worth noting since in Sect. \ref{sec:line} we show stronger evidence supporting a scheme of eDIG ionization with OB stars as the primary source along with other contributing ionization mechanisms. 

We also investigated correlations between the eDIG properties and other global galaxy properties, such as $D_{25}$ (the diameter of the \emph{B}-band isophote corresponding to 25 mag arcsec$^{-2}$; \citealt{vaucouleurs1991rc3}), the scale length ($h_{R}$), and the heights of the stellar disks ($h_{z\text{t}}$ and $h_{z\text{T}}$), $M_*$, and the color excess (the difference between an object's observed color index and its intrinsic color index), but find no clear correlations. However, this does not imply anything definitive about the underlying population of nearby low-mass disk galaxies, as our $M_*$ range is very limited and we lack the necessary statistics.

\section{Comparison with GALEX data}
Ionizing stellar emission comes in the form of LyC photons that have wavelengths below 912 Å. While GALEX FUV does not cover LyC, it is nonetheless a good proxy to map the emission of bright, ionizing stars. FUV traces O and B stars down to ~3 $M_{\odot}$ \citep{kennicutt2012sf}. By comparing the FUV and the H$\alpha$ emission we can infer information about the age of the ionizing stellar population, as hotter early type stars emit more in LyC and as such will cause more gas ionization and more H$\alpha$ emission \citep{hoopes2000hafuv, hoopes2001hafuv, seon2011hafuv}. H$\alpha$ emission traces SF that has occurred within ~20 Myr while FUV traces SF that has occurred within ~100 Myr \citep{kennicutt1998sf}. 

\begin{figure}
  \centering
  \includegraphics[width=0.5\textwidth]{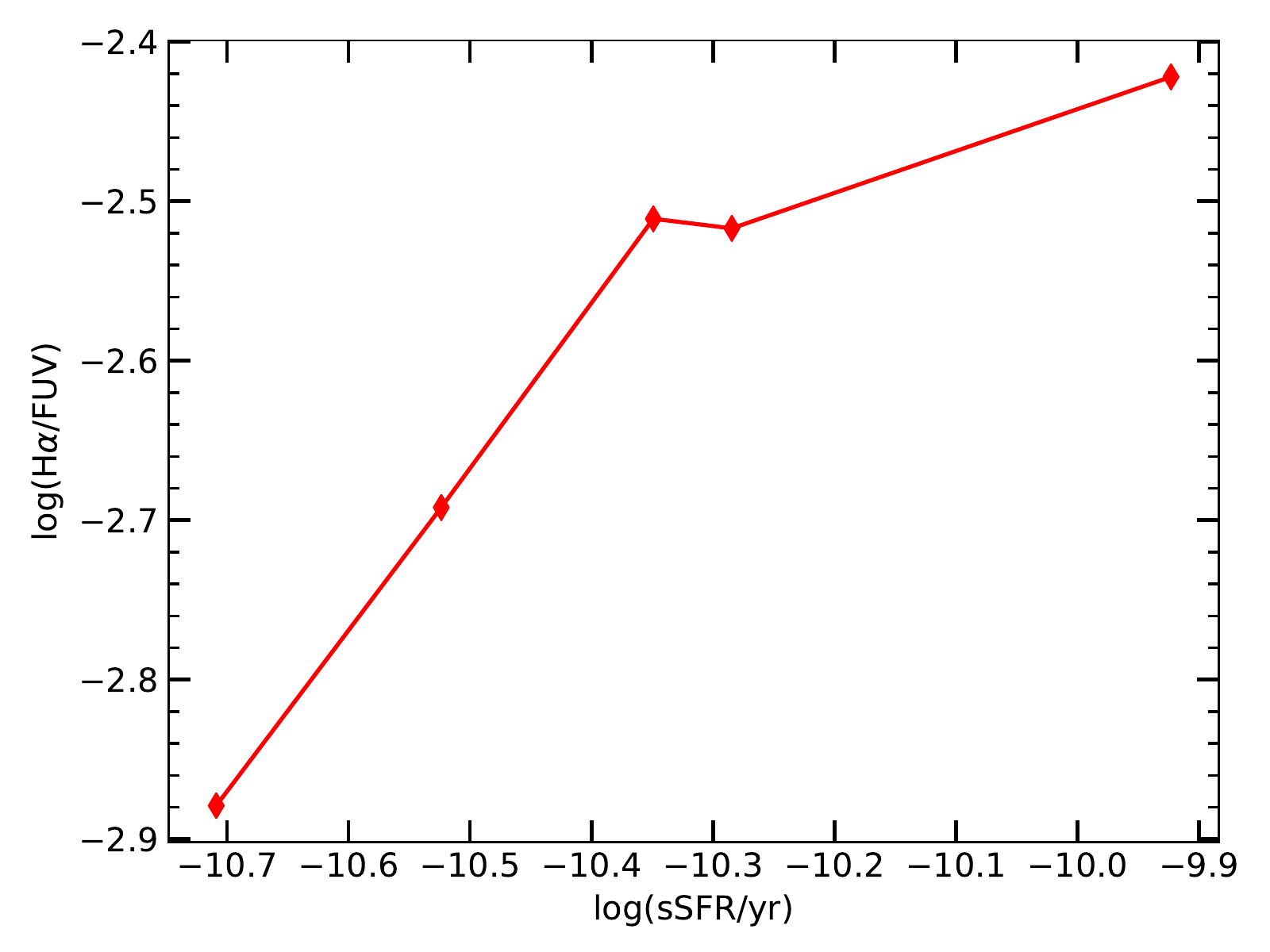}
  \caption{Specific SFR against H$\alpha$/FUV intensity ratio for our sample.}
  \label{fig:hafuv}
\end{figure}

We measured the FUV flux from the GALEX data using the same masks as we did with the H$\alpha$ images. To correct for dust extinction, we again used the Balmer decrement from the MUSE cubes and calculated the FUV extinction adopting a Calzetti extinction law \citep{calzetti2000extinction} with $R_{V}=4.5$ \citep{fischera2005rv}. To gain a measure of the age of the ionizing stellar population, we calculated the H$\alpha$/FUV intensity ratios from extinction-corrected H$\alpha$ and FUV luminosities. The H$\alpha$/FUV intensity ratio correlates well with the H$\alpha$ derived sSFR (Fig. \ref{fig:hafuv}), as one would expect since galaxies with ongoing starbursts naturally have younger stellar populations.

To assess the comparability between the H$\alpha$ and the FUV scale heights, we extracted the vertical FUV profile using the same masking and spatial limits that we used with the vertical H$\alpha$ profiles in Sect. \ref{sec:vprof}. The GALEX data are not sensitive enough for well-constrained double-exponential fits, so we modeled the vertical FUV profile as a single component exponential with a constant background, convolved the model with PSF retrieved from the GALEX online technical documentation\footnote{\url{http://www.galex.caltech.edu/researcher/techdocs.html}}, and fit it to the profile. We obtain FUV scale heights ($h_{z\text{FUV}}$) ranging from 0.26 to 0.98 kpc. The FUV scale heights and extinction-corrected luminosities as well as H$\alpha$/FUV intensity ratios are presented in Table \ref{tab:fuv}.

\begin{table}
  \caption{FUV properties.}
  \label{tab:fuv}
  \centering
  \begin{tabular}{lccc}
    \hline\hline
    ID & $h_{z\text{FUV}}$ & log$(L_{\text{FUV}}/L_{\odot})$ & log(H$\alpha$/FUV) \\
     & (kpc) &  & \\ 
    \hline
    ESO~157-49 & 0.54 & 9.66 & -2.69 \\
    ESO~469-15 & 0.26 & 9.50 & -2.52 \\
    ESO~544-27 & 0.98 & 9.69 & -2.88 \\
    IC~217 & 0.26 & 9.60 & -2.51 \\
    IC~1553 & 0.32 & 10.09 & -2.42  \\
    \hline
  \end{tabular}
  \tablefoot{FUV scale heights ($h_{z\text{FUV}}$) obtained from single-exponential fits. FUV luminosities ($L_{\text{FUV}}$) and H$\alpha$/FUV fluxes (H$\alpha$/FUV) are extinction corrected using the Balmer decrement.}
\end{table}

\begin{figure}
  \centering
  \includegraphics[width=0.5\textwidth]{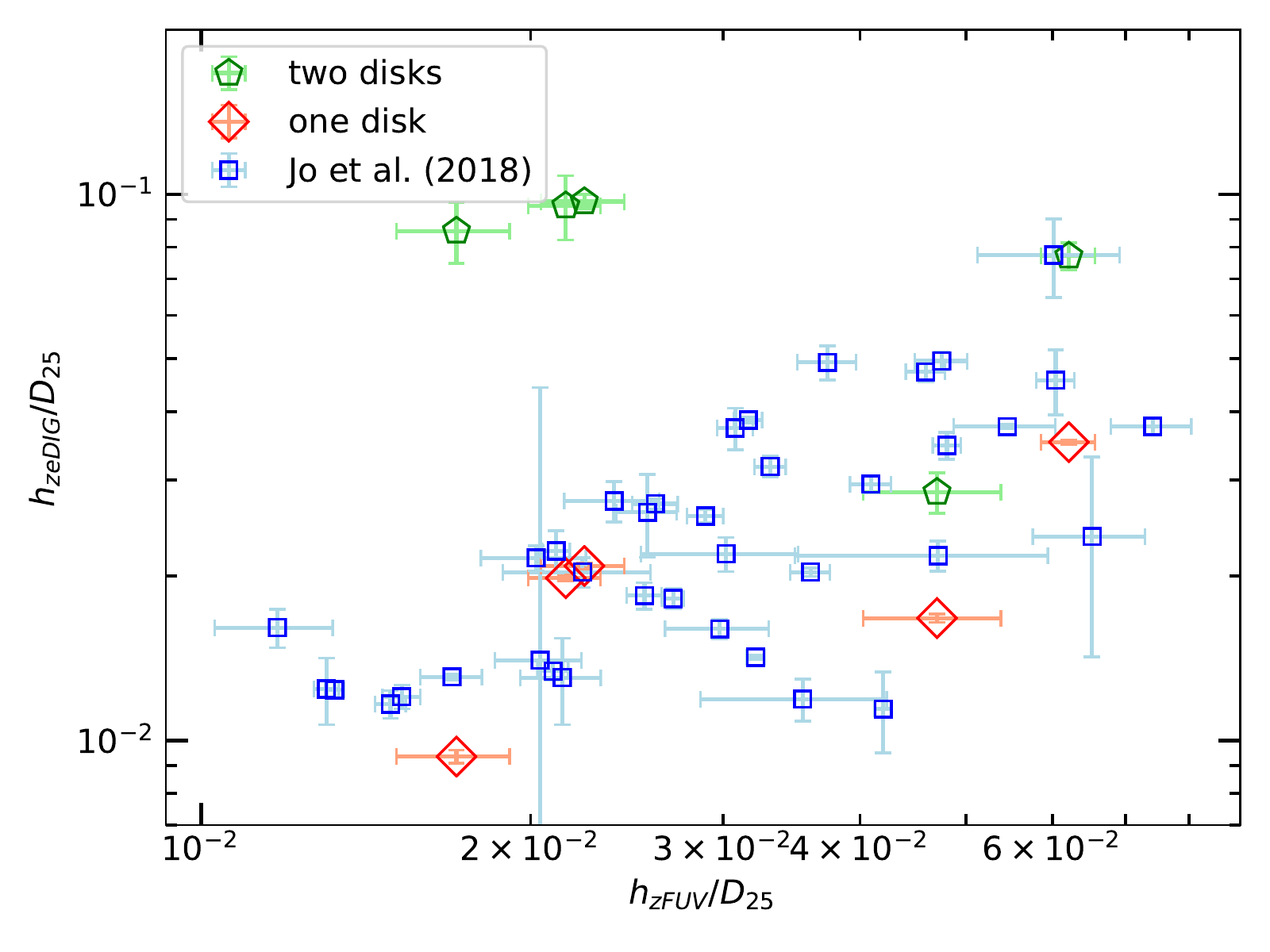}
  \caption{FUV scale height against eDIG scale height for our sample using double-exponential fits (green pentagons) and \cite{jo2018fuv} sample (blue squares). The scale heights are normalized by the $D_{25}$ of the host galaxies. One-standard-deviation error bars for the scale height fits are shown in lighter colors. It should be noted that \cite{jo2018fuv} use single-exponential fits while ours are two-component fits. For comparison, we also show the scale height ratios for our sample using similar single-exponential fits as adopted by Jo et al. (red diamonds).}
  \label{fig:fuvhz}
\end{figure}

\cite{jo2018fuv} find a strong correlation between the H$\alpha$ and the FUV scale heights in their sample of 38 nearby edge-on galaxies using single-exponential fits for both FUV and H$\alpha$. In the galaxies in which they detect extraplanar FUV emission ($h_{z\text{FUV}}$ > 0.4 kpc) they find that the scale height in H$\alpha$ is on average ~0.74 times the scale height in FUV. In contrast, we find that $h_{z\text{eDIG}}$ is larger than $h_{z\text{FUV}}$ for all but one of our galaxies (ESO~544-27). There is also no evidence of any correlation between $h_{z\text{eDIG}}$ and $h_{z\text{FUV}}$ in our sample. However, as can be seen from Fig. \ref{fig:fuvhz}, our single-exponential fit H$\alpha$ scale heights do follow a similar relation with FUV scale heights as the \cite{jo2018fuv} sample. This seems to suggest that the correlation \cite{jo2018fuv} find between the H$\alpha$ and the FUV scale heights may break down at lower surface brightness levels and more thorough two-disk modeling of the eDIG. We however note that our H$\alpha$ imaging is deeper and has better resolution than the GALEX FUV imaging used in \cite{jo2018fuv} as well as in this work. It is possible that with more sensitive FUV data, a two-component morphology might emerge in FUV as well, which might again give rise to a correlation between FUV and eDIG scale heights.

\section{Ionized gas kinematics}

\subsection{Vertical gradients in the ionized gas rotation velocity}
To construct a full picture of the eDIG, we must understand its kinematics in addition to its photometry and morphology. To gain a general sense of the eDIG kinematics in our sample, we constructed position-velocity (PV) diagrams of the ionized gas from the Voronoi binned velocity maps by calculating the vertical and radial distance between the center of the galaxy and the median height ($z$) and radius ($r$) pixel coordinates of each bin. The PV diagrams show evidence of a decrease in the line-of-sight velocity of the ionized gas with increasing distance from the midplane. This is a common property of ionized gas \citep{miller2003lag, bizyaev2017manga, levy2019edge}.

\begin{figure*}[ht]
  \centering
  \includegraphics[width=0.8\textwidth]{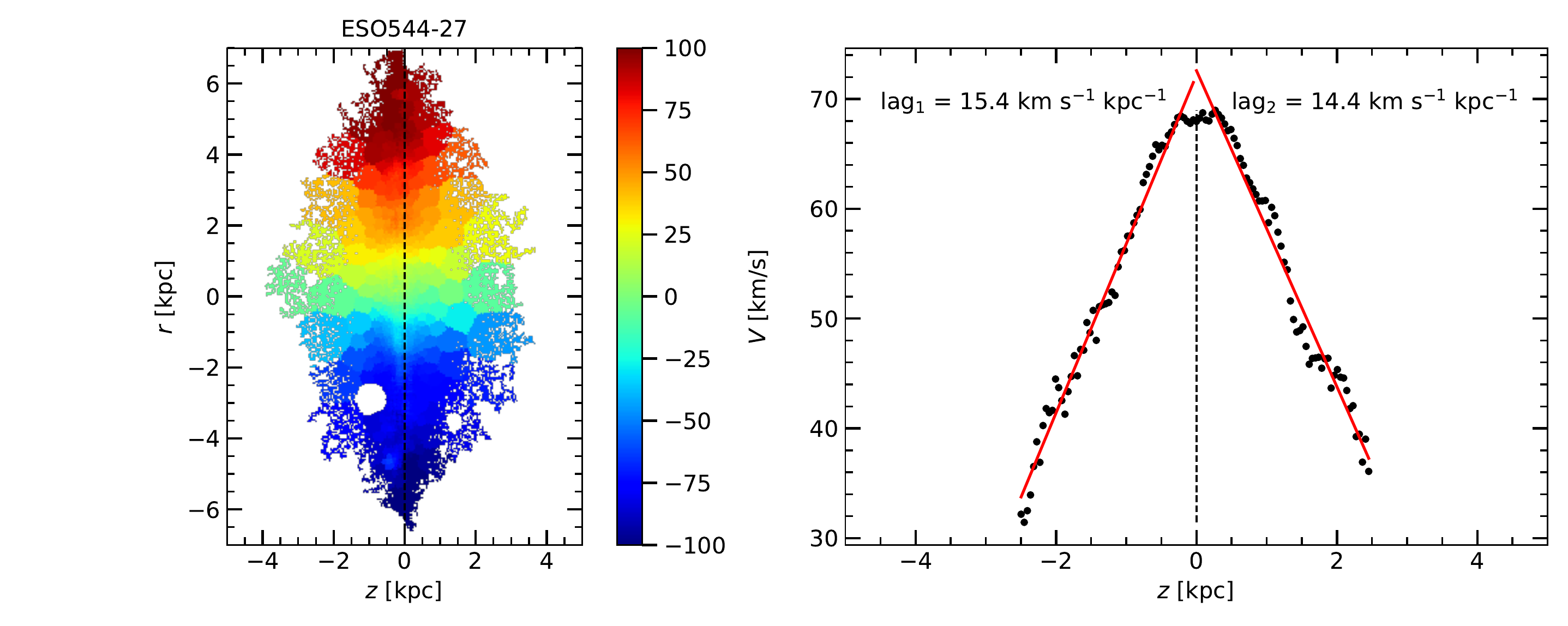}
  \caption{Velocity lag in ESO~544-27. The left panel shows the velocity map, and the right panel shows the vertical velocity profile. The velocity profile has been obtained by averaging the absolute velocity at each height. The black dashed line in both the map and the profile indicates the midplane. The black dots show the measured velocities, and the red lines show the fits to the velocity profile on both sides of the midplane. The absolute values of the slopes of the lines, or the lags above and below the midplane, are given in the right panel.}
  \label{fig:lagex}
\end{figure*}

Such a negative vertical gradient is often called lag, and can be quantified as $-\Delta V / \Delta |z|$ in units of km s$^{-1}$ kpc$^{-1}$. We measured the radially averaged lag from our velocity maps by averaging the absolute velocity in one pixel wide bins at each height above and below the midplane over the disk radial extent to obtain a vertical velocity profile, and then fit a line to this profile above and below the midplane. The slopes of the lines give the lag above and below the midplane. Fig. \ref{fig:lagex} shows the velocity map and vertical velocity profile of ESO~544-27 as an example. The lag (average of the values above and below midplane) is given for each galaxy in Table \ref{tab:kine}. We obtain lag values ranging from 9.76 to 27.4 km s$^{-1}$ kpc$^{-1}$. The velocity maps and PV diagrams are presented in Fig. \ref{fig:lags}.

\cite{levy2019edge} find measurable lags in 60\% of the galaxies in their sample, with values ranging from $-$10 to 70 km s$^{-1}$ kpc$^{-1}$. The median for their complete sample is $19^{+17}_{-26}$ km s$^{-1}$ kpc$^{-1}$, or $21^{+12}_{-27}$ km s$^{-1}$ kpc$^{-1}$ if galaxies with no measurable lag and negative lag (velocity increases with increasing $z$) are excluded. They also compiled lag measurements from literature \citep{miller2003lag, heald2006ngc5775, heald2006ngc891, heald2007ngc4302, bizyaev2017manga} and find a median lag of $25^{+16}_{-28}$ km s$^{-1}$ kpc$^{-1}$. Our lag measurements fall into same range as the \cite{levy2019edge} and other literature values, indicating that the gas kinematics of our sample galaxies are typical of edge-on galaxies.

\begin{table}
  \caption{Ionized gas kinematics.}
  \label{tab:kine}
  \centering
  \begin{tabular}{lcccc}
    \hline\hline
    ID & lag & e$_{\text{lag}}$ & $\partial \text{lag} / \partial \text{r}$ & e$_{\partial \text{lag} / \partial \text{r}}$ \\
     & \multicolumn{2}{c}{(km s$^{-1}$ kpc$^{-1}$)} & \multicolumn{2}{c}{(km s$^{-1}$ kpc$^{-2}$)} \\ 
    \hline
    ESO~157-49 & 12.2 & 2.48 & 13.3 & 4.91\\
    ESO~469-15 & 27.3 & 2.91 & 4.45 & 8.38\\
    ESO~544-27 & 14.9 & 1.71 & 1.62 & 3.18\\
    IC~217 & 9.76 & 1.10 & 0.86 & 6.37 \\
    IC~1553 & 11.5 & 1.60 & 4.91 & 6.70 \\
    \hline
  \end{tabular}
  \tablefoot{The lag is defined as a decrease in rotation velocity with the increase in height, so that positive values correspond to velocity decreasing with height. One-standard-deviation errors (e$_{\text{lag}}$; e$_{\partial \text{lag} / \partial r}$) are also presented for both the average lag and its radial gradient ($\partial \text{lag} / \partial r$).}
\end{table}

\subsection{Radial gradients in the velocity lag}
Different formation scenarios for extraplanar gas can give rise to different radial gradients ($\partial \text{lag} / \partial r$) in the lag \citep{levy2019edge}. For an internal origin of the gas one might expect a decrease in the lag with radius, as for gas ejected to a given height, a centrally concentrated potential will slow down gas more if said gas is ejected from a small radius than if said gas is ejected from a large radius  \citep{zschaechner2015flow}. With accretion that is parallel or inclined to the angular momentum axis, no radial variation in the lag is expected \citep{kaufmann2006accretion,binney2005accretion}. If accretion instead happens from cold gas at the outskirts of the galaxy, the gas will spiral in progressively \citep{combes2014infall}. This could result in larger lags at larger radii if the rotation velocity of the accreted gas increases as it spirals in.

However, regardless of the origin of the gas, the velocity lag and its radial gradient are affected by the shape of the galaxy potential. Therefore, even in the case of gas rotating in circular orbits in a steady axisymmetric galaxy, nonzero velocity lags and radial lag profiles are present.

\cite{levy2019edge} study the radial variations in the lag with the aim of determining the origin of the extraplanar gas. They find that galaxy potential alone is enough to explain the lag radial variations in their sample. They compared $\partial \text{lag} / \partial r$ measured from their sample of 25 edge-on CALIFA galaxies to $\partial \text{lag} / \partial r$ calculated from a Miyamoto-Nagai model potential and find the measured values broadly consistent with the analytic expectation for gas in circular orbits. The model galaxy potential induces lag radial gradients comparable to those they measured. They conclude that the galaxy potential is the primary contributor to $\partial \text{lag} / \partial r$ within their galaxy sample, and the effect of a gas's origins is impossible to discriminate at their resolution and uncertainties.

We attempt a similar analysis for our galaxy sample. Our data may be more amenable to such analysis because of our greater depth and resolution. MUSE has a 1\arcmin $\times$ 1\arcmin FoV sampled at 0\farcs2 $\times$ 0\farcs2, compared to the 331 science fibers covering a FoV of 74\arcsec $\times$ 64\arcsec of the Potsdam Multi Aperture Spectrograph (PMAS; \citealt{roth2005pmas}) in the PMAS fiber package \citep[PPak;][]{verheijen2004ppak, kelz2006ppak} mode used by CALIFA. Additionally, our MUSE data consist of more than two hours of exposure for each galaxy with the 8.2 m VLT, compared to the 45 minutes of exposure per galaxy with the 3.5 m telescope of the Calar Alto observatory in the CALIFA V500 setup \citep{sanchez2012califa}. This allows us to more accurately measure $\partial \text{lag} / \partial r$ and to plot lag as a function of $r$ for individual galaxies.

To take line-of-sight integration effects properly into account, we used an N-body simulation model to estimate the typical $\partial \text{lag} / \partial r$ caused by the potential. The N-body model that we use is the same axisymmetric equilibrium model that is used in \cite{comeron2019thick} to test for retrograde motion in thin$+$thick disk galaxies. Its parameters were chosen to represent low-mass galaxies with no significant bulge component -- similar to our sample galaxies. When applied to our observed galaxies, the N-body data were scaled to have a similar radial stellar disk scale length and maximum rotation velocity as the observed galaxy. More details of the model and how it is constructed are available in \cite{comeron2019thick} and \cite{salo2017barlens}.

\begin{figure*}[ht]
  \centering
  \includegraphics[width=\textwidth]{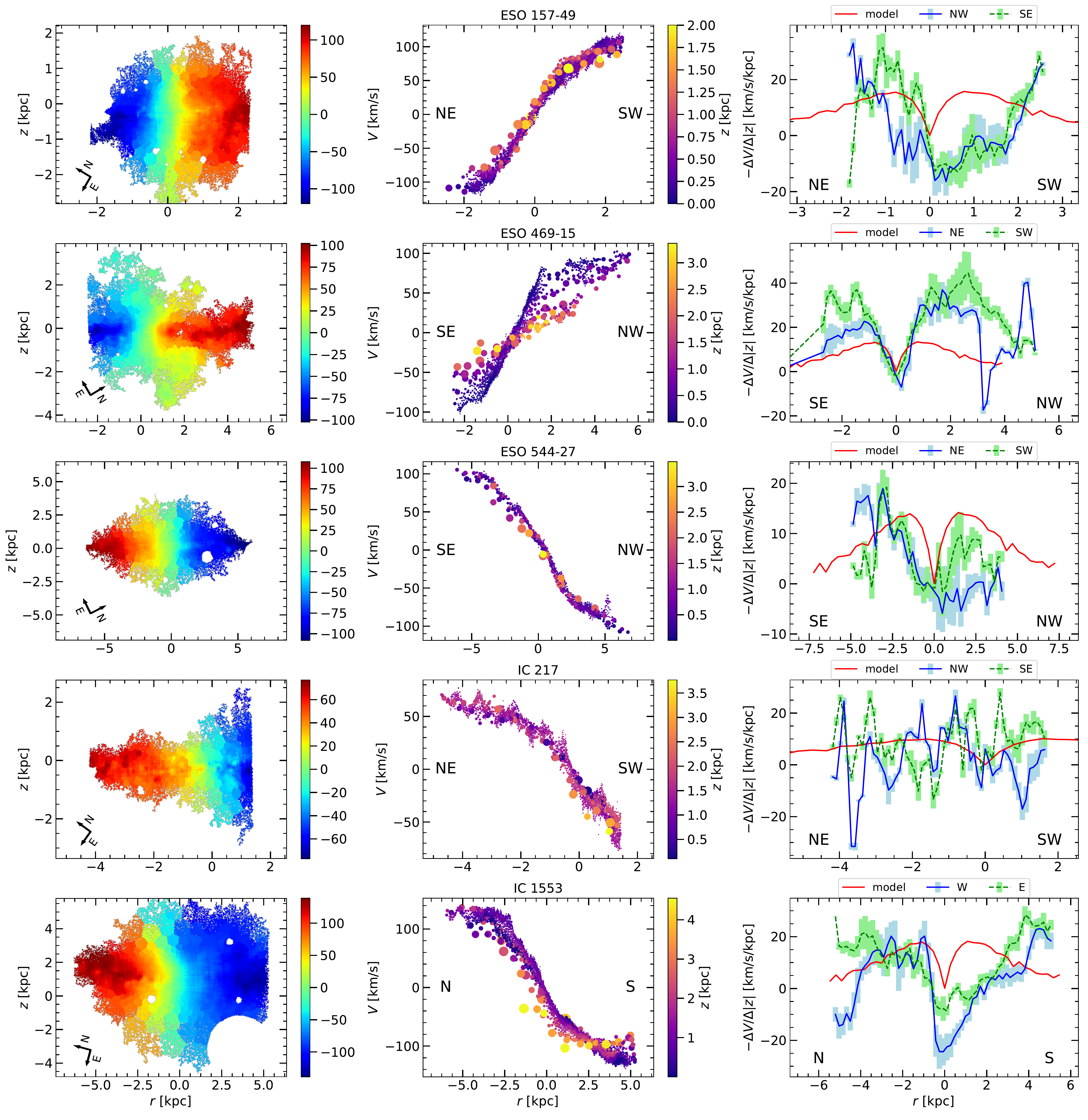}
  \caption{Velocity maps (left), PV diagrams (middle), and lag profiles (right) for the galaxies of our sample. The size of the points in the middle column corresponds to the size of the Voronoi bins they represent. The red curves in the right column show the lag profile of our N-body simulation, scaled to the maximum rotation velocity and to the scale length of the galaxy; this profile corresponds to circular orbits in a steady axisymmetric potential. The solid blue lines and the dashed green lines in the right column show the measured lags above and below the midplane of the galaxy, with the direction indicated in the legend. The uncertainties of the two lag fits are shown in lighter colors. Directions on the sky are indicated in the plots.}
  \label{fig:lags}
\end{figure*}

To obtain the radial profiles of the velocity lag in our sample galaxies, we measured the lag as above, except we did not average over the whole disk and instead measured the lag separately at each radius in 1\arcsec~bins. From this we obtained the radial lag profiles above and below the midplane separately. Similarly, we obtained the potential induced radial lag profile of our N-body simulation. Compared to the Miyamoto-Nagai model derived lag curve of \cite{levy2019edge}, our model curve is qualitatively quite similar. The measured and model lag profiles are compared in Fig. \ref{fig:lags}. We also calculated linear fit $\partial \text{lag} / \partial r$ for our galaxies, which can be found in Table \ref{tab:kine}.

All of the measured lag profiles are highly disturbed compared to the N-body model lag profile. For IC~217 and the southwestern side of ESO~544-27 however, while the measured profiles are disturbed, they more or less follow the model profile and do not deviate from it systematically. ESO~469-15 on the contrary, has significantly stronger lag at all radii compared to the simulation model. This results also in a steeper radial lag gradient in the central parts of the galaxy. This gradient is likely not related to radial inflow but rather to the extraplanar H~{\sc ii} regions of ESO~469-15, which seem to have greater lag than the more diffuse extraplanar gas of the other galaxies. ESO~157-49, IC~1553, and to a lesser extent the northeastern side of ESO~544-27 also have systematic differences to the model. They have highly asymmetric lag profiles, with only one side broadly consistent with the model profile. The opposite sides, by contrast show nearly inverse behavior to that of the model, with consistently increasing lag with radius and negative lag (gas rotation velocity increases with height) near the center of the galaxy. The positive radial gradient of the lag could indicate accretion by radial inflow, although that would not explain the negative lag near the center of the galaxy.

Asymmetries in galaxies are often associated with galaxy interactions or accretion. In our sample only ESO~157-49 has nearby companions; the other galaxies have no neighbors with measured redshifts within 1\degree \ and 1000 km/s. ESO~157-49 has three smaller dwarf neighbors with similar recession velocities, the closest of them being ESO~157-48 located 14.2 kpc projected distance southwest of the galaxy, while the others have projected distances greater than 100 kpc. The direction to ESO~157-48 is parallel to the plane of ESO~157-49, and ESO~157-49 has a positive radial lag gradient on the side nearest to ESO~157-48. The asymmetry and the positive lag gradient of ESO~157-49 could be thus related to interactions with this dwarf companion. As IC~1553 has no known nearby companions, accretion from a diffuse source seems like a more plausible explanation for its distorted eDIG morphology and kinematics.

\section{Emission-line diagnostics}
\label{sec:line}

\subsection{H$\alpha$ equivalent width in the eDIG}
Above we showed that SF in the thin disk correlates well with eDIG ionization. To confirm the significance of OB stars to eDIG ionization and to investigate the contribution of evolved stars and shocks, we examined the gas recombination emission-line spectrum. The equivalent width (EW) of the H$\alpha$ emission line can be used to identify ionization by hot low-mass evolved stars (HOLMES; \citealt{flores-fajardo2011holmes}). HOLMES are post asymptotic giant branch stars and white dwarfs that are plentiful in the thick disks and lower halos of galaxies. Their greater scale height compared to OB stars and their significant contribution to the UV radiation of galaxies \citep{hills1972ism, rose1973uv, terzian1974uv, lyon1975uv}, makes them a promising potential source of eDIG ionization.

The EW of H$\alpha$ is defined as EW(H$\alpha$) = $L_{\text{H}\alpha}/C_{\text{H}\alpha}$, where $C_{\text{H}\alpha}$ is the continuum luminosity around H$\alpha$. As the continuum luminosity is determined mostly by the emission of evolved stars, EW(H$\alpha$) has a strong anticorrelation with the age of the ionizing population \citep{cid2011whan, lacerda2018ew}. The EW(H$\alpha$) distribution among and within galaxies is strongly bimodal \citep{bamford2008ew, cid2011whan, belfiore2016lier, lacerda2018ew}. \cite{lacerda2018ew} interpret the peak at $\sim$3 Å to be caused by a population of HOLMES ionized gas and the peak at $\sim$14 Å to be caused by a population of OB star ionized gas. From this they derive a division of EW(H$\alpha$) < 3 Å as ionization dominated by HOLMES, EW(H$\alpha$) > 14 Å as ionization dominated by OB stars, and 3 Å < EW(H$\alpha$) < 14 Å as a mixed regime where both HOLMES and SF complexes contribute to the ionization. \cite{lopez-coba2019outflows} also use this division, while \cite{levy2019edge} use a simplified scheme where EW(H$\alpha$) = 6 is the division between gas ionized by star-forming H~{\sc ii} regions and gas ionized by other means such as shocks and HOLMES.

We obtained EW(H$\alpha$) maps by dividing the fitted H$\alpha$ flux by the fitted stellar flux at the wavelength of H$\alpha$ in each Voronoi bin. We find no Voronoi bins with EW(H$\alpha$) < 3 Å in any of our galaxies and only few isolated bins with EW(H$\alpha$) < 6 Å. We conclude that HOLMES are not the primary ionization source of eDIG in any of our galaxies. However, we find many extraplanar Voronoi bins that have EW(H$\alpha$) < 14 Å, meaning HOLMES could be a significant secondary ionization source in these regions. The distribution of this secondary HOLMES ionization differs significantly between the galaxies, ESO~544-27 having EW(H$\alpha$) < 14 Å nearly everywhere outside the disk, while IC~1553 has EW(H$\alpha$) < 14 Å only in few extraplanar regions on the northern side of the disk. The EW(H$\alpha$) maps are presented in Fig. \ref{fig:lmaps}.

\begin{figure*}[ht]
  \centering
  \includegraphics[width=\textwidth]{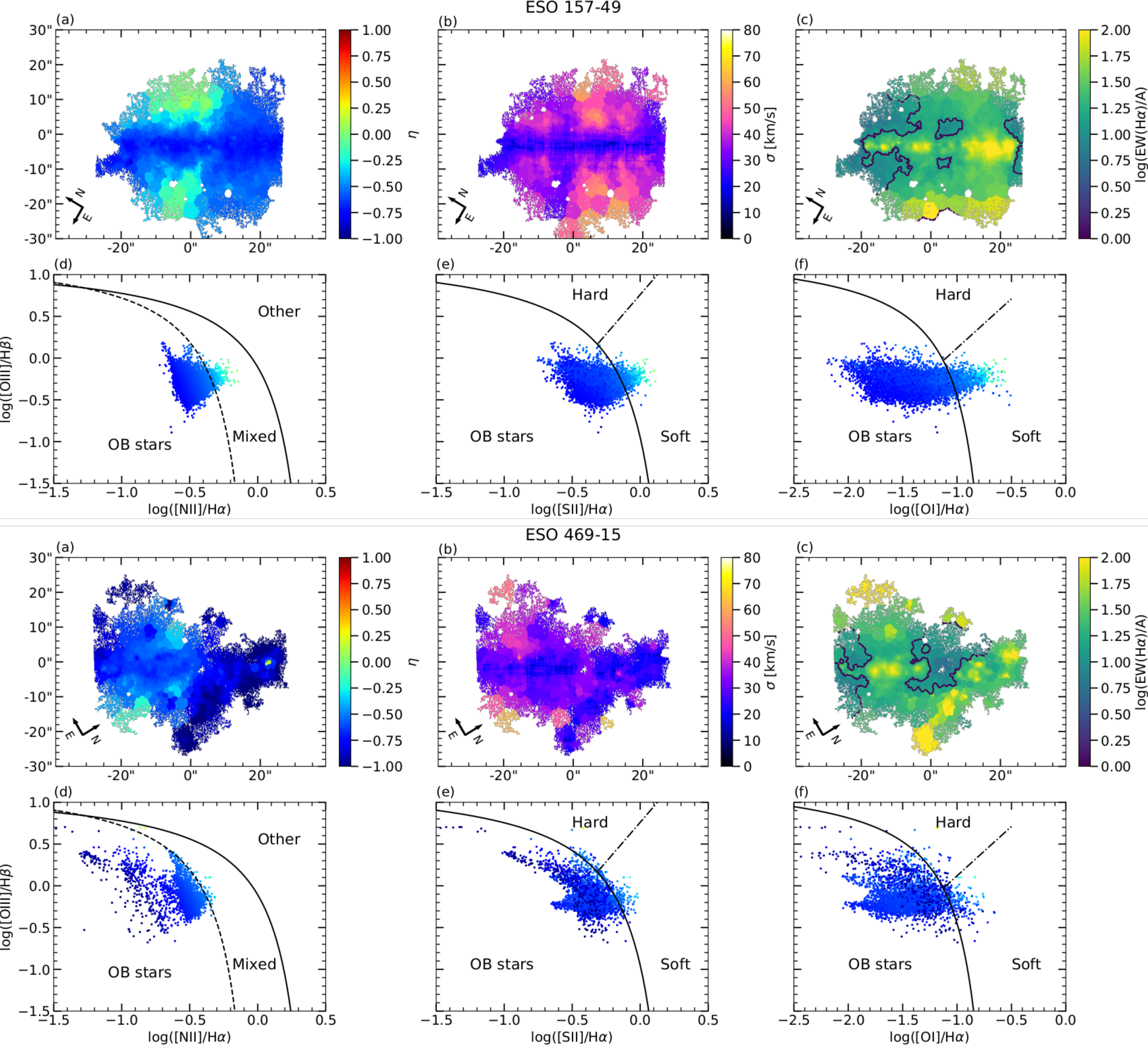}
  \caption{Emission-line properties in our sample galaxies. (a) $\eta$-parameter maps \citep{erroz-ferrer2019eta}. (b) Velocity dispersion maps. (c) Logarithmic EW(H$\alpha$) maps. (d), (e), (f) VO diagrams. The solid line in the VO diagrams is the \cite{kewley2001starburst} extreme starburst line, the dashed line is the \cite{kauffmann2003sfline} empirical starburst line, and the dash-dot line is the \cite{kewley2006agn} Seyfert-LINER demarcation line. The colors in the VO diagrams correspond to the colors in the $\eta$-parameter maps. The lower limit of the OB-star-dominated ionization (EW(H$\alpha$) = 14 Å) is contoured in black on the EW(H$\alpha$) maps.}
  \label{fig:lmaps}
\end{figure*}

\begin{figure*}[ht]
  \centering
  \includegraphics[width=\textwidth]{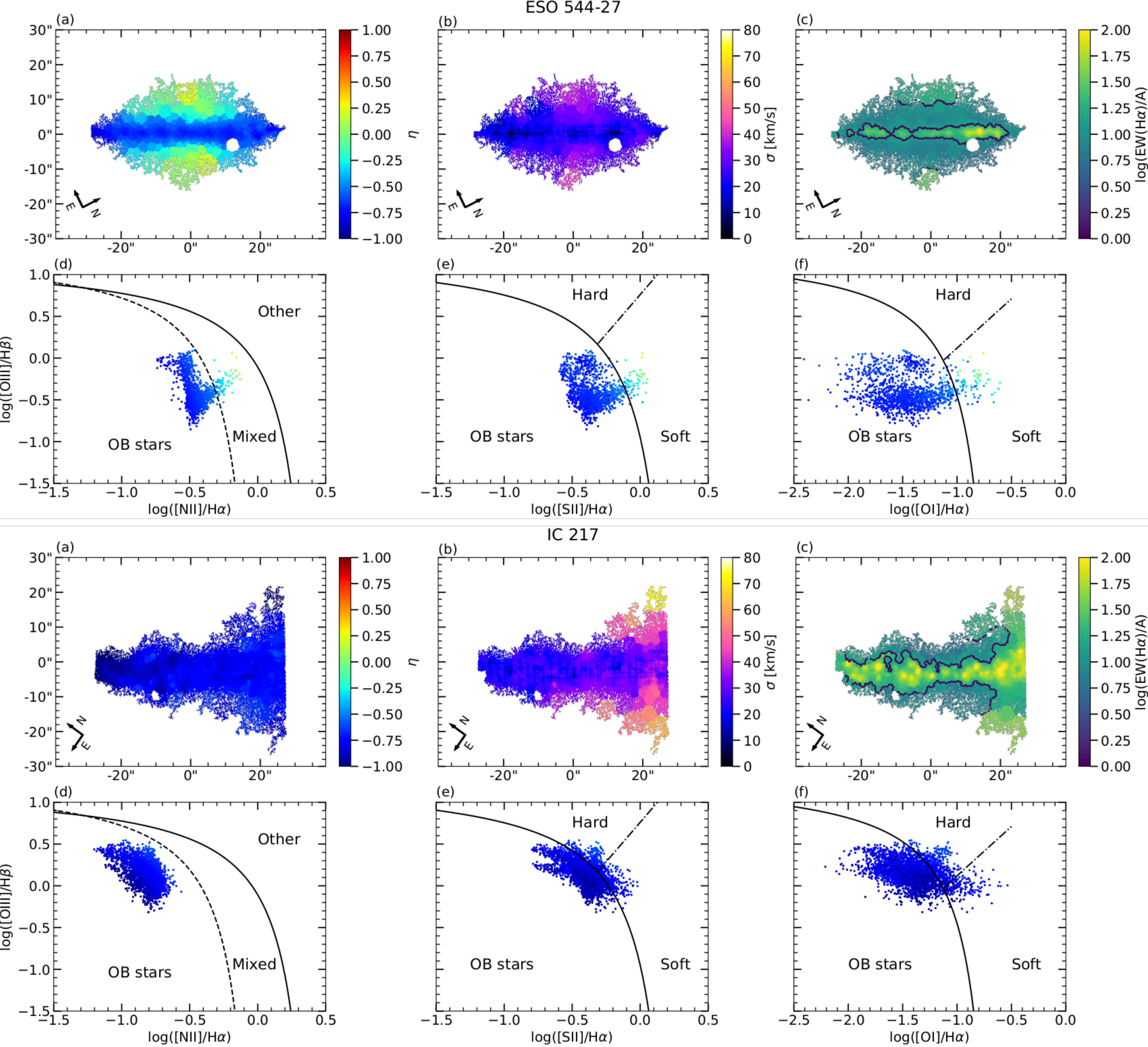}
  \addtocounter{figure}{-1}
  \caption{Continued.}
\end{figure*}

\begin{figure*}[ht]
  \centering
  \includegraphics[width=\textwidth]{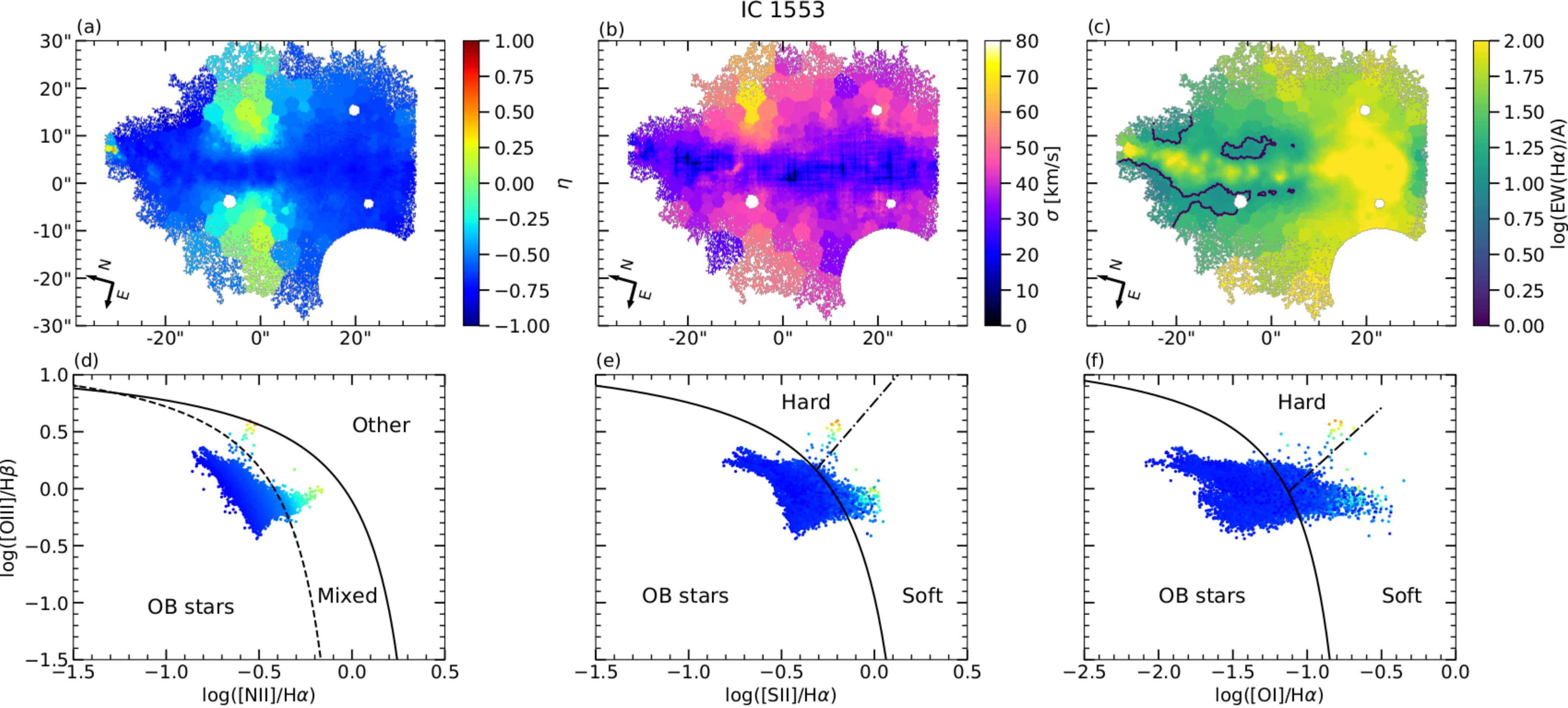}
  \addtocounter{figure}{-1}
  \caption{Continued.}
\end{figure*}

\subsection{Emission-line ratios in the eDIG}
To gain more information about the ionizing spectrum, we constructed emission-line ratio maps from the Voronoi binned MUSE data. We confirm that [N{~\sc ii}]/H$\alpha$, [S{~\sc ii}]/H$\alpha$ and [O{~\sc i}]/H$\alpha$ line ratios are generally enhanced in the eDIG compared to the in-plane H{~\sc ii} regions, with some differences between galaxies. We also find [O{~\sc iii}]/H$\beta$ to be enhanced to some degree in the eDIG of all galaxies except IC~217.

As a single line ratio may depend on many factors such as gas density or temperature, diagnostic diagrams using two different line ratios are better at discriminating different ionizing spectra and their sources. We used the Veilleux-Osterbrock (VO; \citealt{veilleux1987vo}) suite of diagnostic diagrams, EW(H$\alpha$) maps, and ionized gas velocity dispersion ($\sigma$) maps to identify shock ionization in our sample galaxies. Firs, we used the classic [N{~\sc ii}]/H$\alpha$ versus [O{~\sc iii}]/H$\beta$ Baldwin-Phillips-Terlevich (BPT; \citealt{baldwin1981spectra}) diagram, in conjunction with the \cite{kewley2001starburst} extreme starburst line and the \cite{kauffmann2003sfline} empirical starburst line to demarcate regions of OB-star-driven ionization, mixed ionization, and AGN, shock, or HOLMES ionization. We find no ionization dominated by non-OB sources, but we do find many regions with mixed ionization in the BPT diagrams of ESO~157-49, ESO~544-27, IC~1553, and to a lesser extent in the BPT diagram of ESO~469-15. All the ionization in IC~217 falls into the OB star regime on the BPT diagram. We also used the [S{~\sc ii}]/H$\alpha$ versus [O{~\sc iii}]/H$\beta$ and [O{~\sc i}]/H$\alpha$ versus [O{~\sc iii}]/H$\beta$ VO diagrams and the \cite{kewley2006agn} demarcation line separating Seyferts and low-ionization nuclear emission-line regions (LINERs) to further discriminate the ionization between sources with a hard ionizing spectrum, such as AGNs, and a soft ionizing spectrum, such as shocks and HOLMES. We find no evidence of AGNs in our sample galaxies, and we find gas ionized by a hard spectrum only in IC~1553. The [O{~\sc iii}] enhanced emission in IC~1553 originates from a small in-plane region at the northern edge of the disk. It most likely corresponds to some local hard ionizing source such as a Wolf-Rayet star or a supernova.

To better visualize the spatial distribution of the mixed ionization gas in our galaxies and to compare it to the distribution of EW(H$\alpha$), we also constructed $\eta$-parameter maps \citep{erroz-ferrer2019eta}. The $\eta$-parameter is defined as the distance from the bisection of the \cite{kewley2001starburst} and \cite{kauffmann2003sfline} demarcation lines so that $\eta \leq -0.5$ indicates OB star ionization, $\eta \geq +0.5$ AGN, shock, or HOLMES ionization and $-0.5 \leq \eta \leq +0.5$ mixed ionization. Mixed ionization regions that are easily traceable via the $\eta$-parameter in ESO~157-49, ESO~469-15, and IC~1553, are not obviously traced by their EW(H$\alpha$) distributions. In ESO~544-27, by contrast, $\eta$ and EW(H$\alpha$) correlate well for mixed-ionization regions. This suggests that the stellar populations alone can explain the spatial distribution of $\eta$ in ESO~544-27, but additional ionization sources (i.e., shocks) are required to explain its distribution in the other three galaxies. ESO~544-27 also has $h_{z\text{T}} \approx h_{z\text{eDIG}}$, suggesting that HOLMES in thick disk are responsible for its mixed eDIG ionization.

As shock ionized gas produces emission lines with a velocity dispersion around the mean shock velocity, it can be unambiguously identified if an enhanced velocity dispersion is observed in correlation with shock-sensitive emission-line ratios. From purely shock ionized gas, one would expect $\sigma > 80$ km s$^{-1}$, while photoionized gas typically has $\sigma < 40$ km s$^{-1}$ \citep{kewley2019el}. To test our shock ionization interpretation, we obtained velocity dispersion maps from our Voronoi binned MUSE data. We do not find $\sigma > 80$ km s$^{-1}$ in any galaxy of our sample, but this is to be expected as our BPT diagrams point toward mixed ionization regions, rather than fully shock-ionization-dominated regions. We do find that enhanced velocity dispersion correlates well with the $\eta$-parameter in IC~1553, confirming our interpretation that the mixed ionization in IC~1553 is caused by shocks. ESO~544-27 on the other hand has $\sigma < 40$ km s$^{-1}$, consistent with photoionization, confirming that the mixed ionization in ESO~544-27 is caused by HOLMES. ESO~157-49 -- the other galaxy with significant mixed ionization regions -- does have enhanced $\sigma$, but the spatial correlation between the mixed ionization regions and enhanced velocity dispersion is weak. While there is some overlap in the central parts of the galaxy, the mixed ionization regions are concentrated on the northeastern side of the galaxy, while the enhanced velocity dispersion is concentrated on the southwestern side of the galaxy.

The $\eta$-parameter maps, velocity dispersion maps, and the VO diagrams are presented in Fig. \ref{fig:lmaps}. Overall, we find no regions where AGNs, shocks, or HOLMES dominate the ionization in any of our galaxies. This, combined with our morphological H$\alpha$ analysis confirms that radiation by in-plane OB stars is the primary -- although not the only -- ionization source of eDIG in our sample galaxies. We now provide more granular descriptions of the ionization sources in our individual sample galaxies, to showcase the great variation displayed even among this sample of five low-mass disks.

\section{Summary of individual galaxies}
Our sample consists of isolated disk galaxies of similar size, covering only a narrow mass range. Despite this apparent homogeneity, the eDIG properties and structure vary considerably between the galaxies. In the following, we provide qualitative descriptions of eDIG morphologies and summarize the properties of each galaxy to highlight the complexity of structures present in the eDIG.

\subsection{ESO~157-49}
ESO~157-49 is the most compact galaxy in our sample, with the smallest D$_{25}$ and $r_0$. It also has the second smallest $h_{z\text{eDIG}}$ and is the dustiest of our galaxies. Despite its relatively small $h_{z\text{eDIG}}$, it has a rich eDIG morphology with filaments and diffuse emission. One large filament extends up to 3 kpc above the disk on the southeastern side of the galaxy. The eDIG and disk radial H$\alpha$ emission correlation is the second weakest in our sample.

The radial distribution of the vertical gas lag is highly asymmetric and contains negative lags and a rising radial lag gradient on the southwestern side of the galaxy. The disk of the galaxy is also slightly brighter in H$\alpha$ on the southwestern side, although the radial H$\alpha$ profile of the disk is fairly uncertain due to large amount of dust contamination.

The emission-line properties of the extraplanar gas of ESO~157-49 mostly agree with an OB-star-driven ionization, although there are regions corresponding to mixed OB--shock ionization, as well as mixed OB--HOLMES ionization, especially on the northeastern side of the galaxy. There is an enhancement in the velocity dispersion in the eDIG of ESO~157-49, but it does not correlate with enhanced [N{~\sc ii}]/H$\alpha$.

ESO~157-49 is also the only galaxy in our sample with known nearby companions. The closest of them,  the dwarf ESO~157-48 is located at 14.2 kpc in projected distance toward the southwest, in a direction parallel to the plane of ESO~157-49.

\subsection{ESO~469-15}
ESO~469-15 is the least massive galaxy in our sample. It also has the second largest $h_{z\text{eDIG}}$ and second strongest disk-eDIG radial H$\alpha$ profile correlation. In addition to diffuse and filamentary extraplanar ionized gas, it has discrete extraplanar H{~\sc ii} regions. These extraplanar H{~\sc ii} regions are found at up to 3 kpc height above the disk and are also spread further radially than the H$\alpha$ luminous disk, being found up to 8 kpc away from the center of the galaxy, at the edges of the stellar disk. They have [S{~\sc ii}]/H$\alpha$ and [O{~\sc i}]/H$\alpha$ ratios as well as EW(H$\alpha$) consistent with the disk H{~\sc ii} regions, but enhanced [O{~\sc iii}]/H$\beta$ and reduced [N{~\sc ii}]/H$\alpha$ ratios. This implies a harder ionizing spectrum than disk H{~\sc ii} regions.

The ionized gas of ESO~469-15 has a significant vertical lag (more than 30 km s$^{-1}$ kpc$^{-2}$ locally) that cannot be explained by the potential only. This enhanced lag may be caused by the extraplanar H{~\sc ii} regions lagging more than the diffuse gas. This would suggest that the extraplanar H{~\sc ii} regions have different origin than the diffuse gas. Alternatively, this enhanced lag of the extraplanar H{~\sc ii} regions may be related to the slight warp present in the disk of ESO~469-15.

The extraplanar gas of ESO~469-15 is mostly ionized by OB stars. There are also some regions consistent with mixed OB--HOLMES ionization and a few small regions consistent with mixed OB--shock ionization.

\subsection{ESO~544-27}
ESO~544-27 is the second most massive galaxy in our sample, but it has the lowest SFR. It also has the smallest $h_{z\text{eDIG}}$, weakest disk--eDIG radial H$\alpha$ profile correlation and smallest H$\alpha$/FUV intensity ratio. Its $h_{z\text{FUV}}$ is largest in the sample. Its eDIG morphology is quite simple, with only faint filamentary structure.

The radial distribution of lag in ESO~544-27 is roughly consistent with our model potential lag on the southwestern side of the midplane. On the northeastern side, the profile is similarly asymmetric as those of ESO~157-49 and IC~1553. Like ESO~157-49 and IC~1553, the side with rising lag gradients and negative lags has enhanced H$\alpha$ emission.

The eDIG emission in ESO~544-27 falls mostly in the mixed ionization regime in both the EW(H$\alpha$) map and the $\eta$-parameter map. As the velocity dispersion map shows no indication of shocks in ESO~544-27, the emission-line properties of its eDIG are most likely caused by a significant population of HOLMES in its halo. This is also supported by ESO~544-27 being the least active galaxy in our sample and having the oldest stellar population as traced by the H$\alpha$/FUV intensity ratio.

\subsection{IC~217}
IC~217 has a very noticeably asymmetric eDIG morphology. The southwestern side of the galaxy has eDIG filaments that extend to up to 2 kpc from the midplane, while the northeastern side has only minimal eDIG. This asymmetry is also seen in the disk H$\alpha$ brightness, with the southwestern side being more H$\alpha$ luminous. Its FUV disk is also very thin with the smallest $h_{z\text{FUV}}$.

ESO~157-49 and IC~1553, the two other galaxies with asymmetric H$\alpha$ distributions have negative lags and rising radial lag gradients on the side of the stronger H$\alpha$ emission. If those features are indeed related and possibly connected to accretion, one would expect IC~217 to also have negative lags or rising radial lag gradients on its H$\alpha$ bright side. Unfortunately, we cannot confirm this as our MUSE pointing of IC~217 only covers its dimmer northeastern side.

The eDIG ionization of IC~217 is mostly mixed OB--HOLMES ionization. On the southwestern edge of the MUSE pointing the ionization is pure OB star ionization, indicating that the more extended eDIG on the southwestern side may be entirely ionized by OB stars. Again, this cannot be confirmed due to the limits of the MUSE FoV. There is no hint of shock ionization in IC~217. 

\subsection{IC~1553}
IC~1553 is the most massive and most active galaxy in the sample. It has the highest total and sSFR, the highest H$\alpha$/FUV intensity ratio, the largest $h_{z\text{eDIG}}$, and the strongest disk--eDIG radial H$\alpha$ profile correlation. It has a rich eDIG morphology with filaments, a halo and a single extraplanar H~{\sc ii} region at a 4 kpc distance from the midplane on the western side of the galaxy. The morphology is also asymmetric, with more pronounced eDIG emission on the southern side of the galaxy. The H$\alpha$ disk is also very asymmetric, the southern side being considerably brighter than the northern side. Additionally, both the H$\alpha$ disk and eDIG are severely truncated on the southern side of the galaxy.

IC~1553 has similar lag profile as ESO~157-49 and the northeastern side of ESO~544-27, with the H$\alpha$ bright side having negative lags and rising radial lag gradient. The enhanced H$\alpha$ emission on the rising lag side is very prominent. Like ESO~544-27, IC~1553 has no nearby companions with measured redshifts, suggesting that the rising radial lag gradient and enhanced SF is caused by accretion from a diffuse source.

The eDIG in IC~1553 is mostly ionized by OB stars, with additional regions of mixed OB--shock and OB--HOLMES ionization. Like in ESO~157-49, the OB--HOLMES ionization regions are concentrated in the side of the galaxy that is dimmer in H$\alpha$ (northern side for IC~1553), and the OB-shock ionization is biconical. There is also a small region of gas ionized by a harder spectrum in-plane at the northern edge of the disk. This [O{~\sc iii}] enhanced emission most likely corresponds to some local hard ionizing source such as a Wolf-Rayet star or a supernova.

\section{Discussion}

\subsection{Ionization source}
Our results and previous work imply that ionizing radiation from young massive stars is the most important source of ionization for the eDIG (e.g., \citealt{haffner2009leaky, weber2019leaky, levy2019edge}). One of the most difficult to solve problems regarding OB-star-driven ionization is the large distance between the extraplanar gas and ionizing stars in the midplane. If this were a significant physical limitation, one would assume that we would find progressively more gas ionized by other sources at high altitudes. This is not the case for our sample: we do not find any correlation between the height of the gas and its ionization source. This implies that a significant fraction of the ionizing LyC radiation of OB stars can escape the midplane H~{\sc ii} regions and propagate to heights of several kiloparsecs.

Star formation is not the only significant ionization mechanism for eDIG, as can be seen from our results. While often these other mechanisms are not dominant in terms of ionizing power, they may still explain some of the properties of the extraplanar gas that are not easily explained by OB-star-driven ionization, such as [N{~\sc ii}]/H$\alpha$, [S{~\sc ii}]/H$\alpha$, [O{~\sc i}]/H$\alpha$, and [O{~\sc iii}]/H$\beta$ emission-line ratios. To quantify the relative ionizing powers of the different mechanisms, spectral synthesis modeling is required. Doing this in the context of our sample will be the subject of a future paper.

The ionization of the eDIG is not uniform. The mixed ionization that we find is always localized and almost always has a clear source. The best examples of this in our sample are the biconical regions of OB--shock ionization in ESO~157-49 and IC~1553. They are positioned symmetrically on both sides of the midplane, are radially confined, and seemingly originate from the disk. They are likely caused by a large bidirectional outflows related to a superbubbles that have broken out of the disk \citep{norman1989sb}. The starbursts responsible for these superbubbles have quenched more than 20 Myr ago, as there are no corresponding H$\alpha$ emission enhancements in the disk for either of the galaxies. On the contrary, the outflows seem to originate from low-SFR regions, especially for IC~1553, indicating possible feedback in the form of SF quenching by superbubbles \citep{keller2016sb}.

As an older population of stars, HOLMES are much more evenly distributed in the thick disk and stellar halo, compared to the young in-plane OB stars. The regions of mixed OB--HOLMES ionization in EW(H$\alpha$) maps indicate local differences in the ability of the in-plane OB stars to ionize the eDIG. Where OB star ionization is weakened, HOLMES ionization is revealed. This is evident in most of our sample galaxies as the mixed OB--HOLMES ionization regions correlate with reduced disk SF. HOLMES in the thick disk can be an especially important source of eDIG ionization for low-SFR galaxies such as ESO~544-27.

In low-mass disk galaxies eDIG is primarily photoionized, either by OB stars or HOLMES. However, the ionization of the eDIG is clearly not a simple phenomenon explained with a single mechanism. In different galaxies different ionization mechanisms have different weights. Even in our small sample of five outwardly very similar galaxies there is great variety in the properties of the eDIG. Localized shock ionization by superbubble-driven outflows is present in two out of our five galaxies. A suite of different ionization mechanisms is required to fully understand the nature of the ionization of extraplanar gas. Further work is required with a comprehensive analysis of different ionization mechanisms using larger multiwavelength samples, so that the relative importances of the different ionization mechanisms can be linked with other galaxy properties.

\subsection{Origin of the gas}
The origin of the extraplanar gas is more difficult to ascertain than its ionization source. While the emission-line properties of the eDIG can be used to determine the ionization source with relative certainty, kinematics and modeling can only give us hints of the potential gas origins. Making matters more difficult, extraplanar gas is usually kinematically disturbed \citep{wang1997turb}, as is the case for all galaxies in our sample.  In addition to the origin of the gas, the kinematics of the extraplanar gas are also affected by the gravitational potential of the galaxy. The gravitational potential itself depends on the shape, size, and mass distribution of the galaxy, as determined by its formation history. As such, both the gas origin and the galaxy potential are linked to the evolutionary history of the galaxy, making disentangling their effects on the kinematics very difficult.

As can be seen from the lag profiles of our galaxies (Fig.~\ref{fig:lags}), the eDIG is highly disturbed and neither a steady galaxy model such as our N-body simulation nor a simplistic accretion or galactic fountain model that predicts uniformly rising or decreasing radial lag gradient can explain its observed kinematics. Simulations taking into account the hydrodynamics of the gas are required to model the complex observed trends.

Regardless of the difficulties involved with the kinematic analysis, we find hints pointing toward an accretion origin for the extraplanar gas in four out of our five galaxies. The asymmetry displayed by the ionized gas morphology in IC~217 and IC~1553 could be explained by accretion, especially considering that galaxy interactions are unlikely to be the cause of the asymmetry due to the lack of nearby companions with known redshifts for all of our galaxies except ESO~157-49. The rising lag gradients present in ESO~157-49, ESO~544-27, and IC~1553 could also indicate accretion. The spatial correlation with rising lag gradients and increased SF in the disk present in ESO~157-49, ESO~544-27, and IC~1553 is also expected, as accretion is known to be a trigger of SF \citep{binney2005accretion, sancisi2008accretion}. The galaxies that have the strongest evidence of accretion (ESO~157-49 and IC~1553) also have the largest in-plane DIG scale heights ($h_{z\text{DIG}}$) and smallest stellar thick disk scale heights ($hz_{\text{T}}$), possibly suggesting enhanced global SFR and SF scale height due to gas accretion.

A spatial correlation between the eDIG morphology and the disk SF would also be expected with galactic fountain origin of the gas. In IC~217 and IC~1553 the eDIG is noticeably more extended near the regions of high SF in the disk. Discriminating between internal and external origin becomes even more difficult when considering that galactic fountains can drive accretion \citep{combes2014infall, fraternali2017both}. Most likely, much like with the ionization source, a composite model is required to fully explain the origin of the eDIG.

\section{Summary and conclusions}
We have investigated the eDIG properties of five low-mass edge-on galaxies using MUSE observations, deep H$\alpha$ narrowband imaging obtained with the NTT and the NOT, and archival GALEX FUV data. We have performed photometric, kinematic, and abundance analyses of the eDIG with the aim of determining the origin and ionization source of the extraplanar gas. We summarize our main results below.
\begin{enumerate}[]
\item We measured the eDIG scale heights using two-component exponential fits corrected for the extended PSF. We find $h_{z\text{DIG}} = $ 0.12--0.24 kpc for the thinner quiescent component and $h_{z\text{eDIG}} = $ 0.59--1.39 kpc for the disturbed extraplanar component. These values are consistent with previous measurements of eDIG scale heights in low-mass galaxies.
\item We calculated the Pearson correlation coefficient of the radial H$\alpha$ profiles of the disk and the eDIG and find a tight correlation for all of our galaxies ($r =$ 0.55--0.94). We interpret this as evidence of the importance of midplane SF to the ionization of eDIG.
\item We find a positive correlation between the sSFR and $h_{z\text{eDIG}}$ and between the sSFR and disk--eDIG Pearson \emph{r} values. This means that the galaxies with the higher sSFR have a more extended eDIG and show a stronger correlation between the in-plane SF and eDIG, further linking SF to the ionization of eDIG.
\item We measured the vertical velocity gradients, or lags, of the ionized gas. For all of our galaxies, the velocity decreases with \emph{z}, with values of lag ranging from $9.8\pm1.1$ to $27.4\pm2.9$ km s$^{-1}$ kpc$^{-1}$. This is consistent with previous measurements in the literature.
\item We investigated the radial variations in the lag. We compared measured radial lag profiles to a profile derived from the potential of an N-body simulated galaxy. All of our galaxies have highly disturbed lag profiles that do not fit with the steady galaxy model, nor can they be explained with the simplistic accretion or galactic fountain models. That said, the gas kinematics in ESO~157-49, IC~1553, and to a lesser extent ESO~544-27 are broadly compatible with an accretion origin for part of the extraplanar gas.
\item We constructed H$\alpha$ EW maps to identify ionization caused by HOLMES. We find no regions in which ionization by HOLMES dominates, defined by EW(H$\alpha$) < 3 Å. We do find localized regions with mixed OB--HOLMES ionization, defined by 3 Å < EW(H$\alpha$) < 14 Å, in all of our galaxies. These regions correlate with a low SFR in the disk.
\item We constructed VO diagrams, $\eta$-parameter maps, and velocity dispersion maps to identify shock ionization. We find regions with mixed OB--shock ionization, including bidirectional outflows most likely caused by superbubble outbreaks, in ESO~157-49, IC~1553, and to a lesser extent in ESO~469-15. These mixed ionization regions are highly localized compared to the photoionization-dominated regions.
\end{enumerate}

The eDIG in low-mass disk galaxies is primarily photoionized, either by OB stars or UV bright evolved stars, while shock ionization by superbubble feedback is always localized and relatively rare. The eDIG is most likely at least partially accreted from the intergalactic or the circumgalactic medium. However, to fully explain the ionization and origin of the eDIG, composite models taking many different mechanisms  into account are required. Further work with larger, more complete samples with good depth, resolution, and wavelength coverage, as well as spectral synthesis modeling and hydrodynamical simulations, is necessary. Linking the importance of the different ionization and formation mechanisms of the eDIG to other galaxy properties, such as SF, stellar mass, and morphology, would give us great insight into feedback mechanisms across galaxy properties. Understanding this liminal gas that separates the domains of the disk and the halo, and is a critical part of their connection and interplay, is vital in understanding baryonic feedback and galaxy evolution.

\begin{acknowledgements}
  We thank the referee on insightful comments that helped to quantify the goodness of our vertical profile fits.
  This research is based on observations collected at the European Southern Observatory under ESO programmes 096.B-0054(A), 097.B-0041(A), and 0102.B-0580(A); as well as on observations made with the Nordic Optical Telescope, owned in collaboration by the University of Turku and Aarhus University, and operated jointly by Aarhus University, the University of Turku and the University of Oslo, representing Denmark, Finland and Norway, the University of Iceland and Stockholm University at the Observatorio del Roque de los Muchachos, La Palma, Spain, of the Instituto de Astrofisica de Canarias; and on observations made with the Galaxy Evolution Explorer, obtained from the MAST data archive at the Space Telescope Science Institute, which is operated by the Association of Universities for Research in Astronomy, Inc., under NASA contract NAS 5–26555.
  RR acknowledges funding from the Technology and Natural Sciences Doctoral Program (TNS-DP) of the University of Oulu.
  HS and AW acknowledge funding from the Academy of Finland grant n:o 297738.
  AW additionally acknowledges support from the STFC [ST/S00615X/1].
  SC acknowledges funding from the State Research Agency (AEI-MCINN) of the Spanish Ministry of Science and Innovation under the grants “The structure and evolution of galaxies and their central regions” with reference PID2019-105602GB-I00/10.13039/501100011033, and  “Thick discs, relics of the infancy of galaxies" with reference PID2020-113213GA-I00.
 SDG acknowledges funding from the European Union’s Horizon 2020 research and innovation programme under the Marie Skłodowska-Curie grant agreement No 893673. SDG additionally acknowledges financial support from the European Union’s Horizon 2020 research and innovation programme under Marie Skłodowska-Curie grant agreement No 721463 to the SUNDIAL ITN network, from the State Research Agency (AEI-MCINN) of the Spanish Ministry of Science and Innovation under the grant "The structure and evolution of galaxies and their central regions" with reference PID2019-105602GB-I00/10.13039/501100011033, and from IAC project P/300724, financed by the Ministry of Science and Innovation, through the State Budget and by the Canary Islands Department of Economy, Knowledge and Employment, through the Regional Budget of the Autonomous Community.
 We acknowledge the usage of the HyperLeda database (http://leda.univ-lyon1.fr), and the NASA/IPAC Extragalactic Database (NED) which is operated by the Jet Propulsion Laboratory, California Institute of Technology, under contract with the National Aeronautics and Space Administration.
 This research made use of python (\url{http://www.python.org}), IRAF, IDL (\url{http://www.harrisgeospatial.com/docs/using_idl_home.html}), SciPy \citep{2020SciPy}, NumPy \citep{2020NumPy}, Matplotlib \citep{matplotlib}, NoiseChisel \citep{gnuastro,noisechisel}, and Astropy, a community-developed core Python package for Astronomy. This work was partly done using GNU Astronomy Utilities (Gnuastro, ascl.net/1801.009) version 0.13.22-69bae. Work on Gnuastro has been funded by the Japanese Ministry of Education, Culture, Sports, Science, and Technology (MEXT) scholarship and its Grant-in-Aid for Scientific Research (21244012, 24253003), the European Research Council (ERC) advanced grant 339659-MUSICOS, European Union’s Horizon 2020 research and innovation programme under Marie Sklodowska-Curie grant agreement No 721463 to the SUNDIAL ITN, and from the Spanish Ministry of Economy and Competitiveness (MINECO) under grant number AYA2016-76219-P. 
\end{acknowledgements}

\bibliographystyle{aa}
\bibliography{ref}{}

\end{document}